\documentclass[final,3p,times,authoryear]{article}
\usepackage{textcomp}
\usepackage{titlesec}

\usepackage{chngcntr}
\titleformat{\chapter}[block]
{\normalfont\huge\bfseries}{\thechapter.}{1em}{\Huge}
\titlespacing*{\chapter}{0pt}{-19pt}{0pt}
\usepackage[margin=3cm]{geometry}
\usepackage{multirow}
\usepackage{wrapfig}
\usepackage{multicol}
\usepackage[usenames,dvipsnames]{color}
\usepackage{wrapfig,lipsum,booktabs}
\usepackage{listings}
\usepackage[section]{placeins}
\usepackage{booktabs}
\usepackage[symbol]{footmisc}
\usepackage{url}
\usepackage[round]{natbib}
\usepackage{float}
\definecolor{light-gray}{gray}{0.9}
\usepackage{color}
\definecolor{yellow2}{RGB}{255,255,255} 
\usepackage{ dsfont }
\usepackage{dcolumn}
\usepackage[utf8x]{inputenc}
\usepackage[T1]{fontenc} 	
\usepackage{graphicx}
\usepackage{caption}
\usepackage{subcaption}
\usepackage{hyperref}
\usepackage{fancyvrb} 
\usepackage[symbol]{footmisc}
\usepackage{listings} 
\usepackage{bm} 
\usepackage{xcolor} 
\usepackage[english]{babel}
\usepackage{graphicx}       						
\usepackage{amsmath,amssymb} 						
\usepackage{siunitx}							      
\usepackage{listings,relsize}
\newcommand{\lil}[1]{\lstinline|#1|}
\title{A geostatistical two field model that combines point observations and nested areal observations, and quantifies long-term spatial variability  -  A case study of annual runoff predictions in the Voss area.}
\author{$\textrm{Thea Roksvåg}^{1}$\footnote{\texttt{thea.roksvag@ntnu.no} - The project is funded by The Research Council of Norway, grant number: 250362.},\hspace{2mm} $\textrm{Ingelin Steinsland}^{1}$ and $\textrm{Kolbjørn Engeland}^{2}$\\
	$^1$ Department of Mathematical Sciences, \textsc{ntnu}, Norway\\
	$^2$The Norwegian Water Resources and Energy Directorate (\textsc{nve})\\}
\begin{document}
\maketitle
	\begin{abstract}
We estimate annual runoff by using a Bayesian geostatistical model for interpolation of hydrological data of different spatial support. That is, streamflow observations from catchments (areal data), and precipitation and evaporation data (point data). The model contains one climatic spatial effect that is common for all years under study, and one year specific spatial effect. Hence, the framework enables a quantification of the spatial variability caused by long-term weather patterns and processes. This can contribute to a better understanding of biases and uncertainties in environmental modeling.

The suggested model is tested by predicting 10 years of annual runoff for around Voss in Norway and through a simulation study. We find that on average we benefit from combining point and areal data compared to using only one of the data types, and that the interaction between nested areal data and point data gives a spatial model that takes us beyond smoothing. Another finding is that when climatic effects dominate over annual effects, systematic under- and overestimation of runoff over time can be expected. On the other hand, a dominating climatic spatial effect implies that short records of runoff from an otherwise ungauged catchment can lead to large improvements in the predictions.
\end{abstract}
	
\section{Introduction}
Data related to meteorology, geology and hydrology are often connected to geographical locations. The data are typically linked to point locations, but there are also data observed over an areal unit, e.g. over a crop field, a forest, a grid from a satellite observation or an administrative unit like a country. While point referenced data give information about the process of interest at one specific location, the areal referenced data impose a constraint on the process and/or contain information about aggregated or mean values in a larger area.
	
For some processes, there exist point data \textit{and} areal data that give information about the same underlying process, and studies show that both observation types should be taken into account when making statistical inference and predictions \citep{areadata1,areadata2}. There are several challenges connected to simultaneously use data of different spatial support: The data types must be connected to the process of interest in a meaningful way, and expert opinions about the involved measurement uncertainties should be taken into account. In addition, information about how the point and areal data are related to each other is important, such that the observation types can be combined in a mathematically consistent way that preserves basic physical laws (i.e. the conservation of mass and energy).
	
In this article we consider runoff, which is an example of a process that can be observed through point and areal data. Runoff is defined as the part of the precipitation that flows towards a river on the ground surface (surface runoff) or within the soil (subsurface runoff or interflow) \citep{WMO}. Every point in the landscape contributes to runoff generation, and on an annual scale runoff can be approximated by the estimated point precipitation minus the actual point evaporation at a location of interest \citep{Sauquet2000}. With this interpretation, runoff is a continuous point referenced process in space. However, runoff accumulated over an area is typically observed by measuring the amount of water that flows through the outlet of a stream. The observed value does not primarily provide information about the runoff at the location of the stream outlet: It primarily provides information about the runoff generating process in the whole drainage area which is called a catchment. Such observations of runoff are therefore areal referenced.
	
Since most catchments in the world are ungauged (i.e. without runoff observations), a common task for hydrologists is to predict runoff in these catchments. In this article we consider predictions of $annual$ runoff which is a key hydrological signature. The annual runoff gives information about the total amount of water available in an area of interest and is fundamental for water resources management, i.e. in the planning of domestic, agricultural, and industrial water supply, and for allocation of water between stakeholders. Annual runoff is also commonly used as a key variable when predicting other runoff properties in ungauged catchments, i.e. low flows and floods \citep{PUB2}. Furthermore, the variability in annual runoff is interesting as it is a key quantity for understanding runoff's sensitivity to driving climatic factors in today's climate, and can be used to make inference about the runoff variability also for future climates. 

There are several approaches to predict runoff in ungauged catchments in hydrology, e.g. process-based methods \citep{Beldring,wasmod} and geostatistical methods \citep{Gottschalk1993b, Sauquet2000,topkriging}. In this article, we choose a geostatistical approach. Within the geostatistical framework, runoff predictions in ungauged catchments have typically been done by interpolation of areal referenced runoff data by using Kriging methods (see e.g.  \cite{topkriging} or \cite{Sauquet2000}). This has shown promising results. In these methods, precipitation data have often been avoided as an information source because these data are known to be uncertain and/or biased (see e.g. \cite{raingauge}, \cite{Groisman} or \cite{Regnusikkerhet}). Evaporation data are even more uncertain: It is seldom observed directly, but derived from meteorological observations and process-based models like in e.g.  \cite{evap1} or \cite{evap3}. In spite of the large uncertainties linked to precipitation and evaporation measurements, precipitation and evaporation are the main drivers behind runoff, and it is reasonable to believe that these data sources can contribute to an increased understanding of the runoff generating process if used cleverly. Particularly in areas with few streamflow observations.

Motivated by this, we present a Bayesian geostatistical model for annual runoff where we in addition to runoff data, use precipitation and evaporation data for spatial interpolation. The suggested model is a Bayesian hierarchical model where the observation likelihood consists of areal referenced runoff observations from catchments and/or point observations of runoff, where the point observations are annual evaporation subtracted from annual precipitation. Informative priors based on expert knowledge are used on the measurement uncertainties to express our doubt on the precipitation and evaporation data, and to put more weight on the runoff observations that are considered more reliable.
	
The catchments we study in this article are located around Voss in western Norway. Voss is a mountainous area, and the areas west for Voss are among the wettest in Europe with annual precipitation around 3 m/year. This makes Voss flood exposed, and accurate runoff models are of high importance. Voss is also a challenging area when it comes to runoff estimation due to large spatial variability and low stream gauge density. However, there are several precipitation gauges in the area that can be exploited to increase the hydrological understanding. This makes the Voss area  a good candidate for performing spatial interpolation of runoff by also including precipitation and evaporation data.
	
The large annual precipitation in western Norway is mainly caused by the orographic enhancement of frontal precipitation formed around extratropical cyclones. The orographic enhancement is explained by steep mountains that create a topographic barrier for the western wind belt, which transports moist air across the North Atlantic \citep{Stohl}. The topography and the elevation differences result in prominent patterns in precipitation and runoff.
	
Motivated by the strong orographic effect, we include a spatial component in the model that is constant over the years for which we have runoff observations. This represents the spatial variability of runoff caused by climatic conditions in the study area.  Furthermore, it is reasonable to assume that not all of the spatial variability can be explained by the climate, and we include an additional spatial effect to describe the annual discrepancy from the climate. 
	
The climatic part of the model is interesting because it let us quantify how much of the spatial variability that can be explained by long-term effects. Separating long-term spatial variability from year dependent effects can lead to a better understanding of systematic biases and uncertainties that occur in the prediction of environmental variables due to weather patterns and processes that are more or less apparent each year. A consequence of including the climatic component is also that we obtain a model for which it is possible to exploit short records of data: The climatic component captures how the short records vary relatively to longer data series from nearby catchments. This is a valuable property because sparse datasets are common in hydrology. There are several studies on how short records of runoff can be used to estimate different hydrological signatures \citep{shortrec1,shortrec2}, but our framework represents a new approach by incorporating the short records into a geostatistical framework where several years of runoff are modeled simultaneously through a climatic spatial field.
	
Making inference and predictions with geostatistical models often lead to computational challenges due to matrix operations on (dense) covariance matrices, and in our suggested model we have not only one, but two spatial fields. Our solution to the computation challenges is to use the \textsc{spde}-approach to spatial modeling from \cite{mainINLA}. \cite{mainINLA} utilizes that a Gaussian random field (\textsc{grf}) with a Matérn covariance function can be expressed as the solution of a stochastic partial differential equation (\textsc{spde}). By approximating the solution of the \textsc{spde} by using the finite element method \citep{FEM}, the involved GRFs can be expressed as Gaussian Markov random fields (\textsc{gmrf}s). The \textsc{gmrf} approximations enable fast simulation and inference \citep{GMRFbook}, and integrated nested Laplace approximations (\textsc{inla}) can be applied \citep{mainINLA}.	
		
In geostatistical methods used for runoff interpolation it is common to link the involved catchments to point locations in space, not to areas (see e.g. \cite{euclid2} or \cite{euclid3}). However, interpreting catchment runoff as point referenced can lead to a violation of basic conservation laws: A significant property of catchments is that they are organized into subcatchments, and for annual runoff the water balance must be conserved for all subcatchments. That is, the total amount of annual runoff in a subcatchment cannot be larger than the total annual runoff in the main catchment. In the Top-Kriging approach developed by \cite{topkriging} the nested structure of catchments is taken into account by computing the covariance between two catchments based on the pairwise distance between all the grid nodes in a discretization of the target catchments. This way, information from a subcatchment is weighted more than information from a nearby non-overlapping catchment. The Top-Kriging approach is currently one of the leading interpolation methods for runoff, and has outperformed other methods in predicting several hydrological signatures in Austria \citep{Austriacompare}. 
	 
Our model is similar to the Top-Kriging approach by that we consider streamflow observations as areal referenced and compute the covariance between two catchments accordingly. However, our methodology differs from Top-Kriging and other hydrological interpolation methods by using precipitation (point) data in the interpolation framework in addition to nested streamflow (areal) data. As this is an important difference, one of the main objectives of this paper is to:  \\
1) Explore how the runoff predictions in Voss are influenced by the two different observation types (point and areal observations), and assess if the combination of point and areal data can contribute to an increased predictive performance. \\\vspace{-3mm}

Furthermore, the model we suggest ensures that the water balance is preserved for any point in the landscape by defining annual runoff in a catchment as the integral of the point runoff over the catchment's area. Top-Kriging and other geostatistical models don't necessarily provide a full preservation of the water balance. A second objective is therefore to: \\
2) Show by example how the interaction between point observations and nested areal observations can contribute to improved predictions of annual runoff because the water balance is taken into account.\\ \vspace{-3mm}

A geostatistical model that combines point and areal data in the same way as we do already exists in the literature in \cite{areadata1}. What is new in our model in terms of statistical modeling is the climatic spatial component.  A final objective of the paper is thus to:\\
3) Present a model for which the spatial variability due to long-term spatial patterns can be quantified, and show how this can be used as a tool for understanding the uncertainty and biases in the modeling of environmental variables, and for exploiting short records of data.

In the section that follows, we present the study area and the available data. Next, we introduce the theoretical background needed to develop the suggested runoff model that is presented in Section \ref{sec:models}. In Section \ref{sec:casestudy} the suggested model is fitted to the Voss data. Based on some observation schemes described in Section \ref{sec:eval}, the predictability of annual runoff in Voss is evaluated and discussed. To further demonstrate the value of including a climatic spatial field in the model, a simulation study was carried out. This is presented in Section \ref{sec:simulationstudy}. Finally, our key findings are discussed in Section \ref{sec:discussion}.

\section{Study area and data}\label{sec:data}
\begin{figure*}[h!!]
	\centering
	\begin{subfigure}[b]{0.45\textwidth}
		\includegraphics[page=1, trim = 0mm 0mm 40mm 0mm, clip, width=6cm]{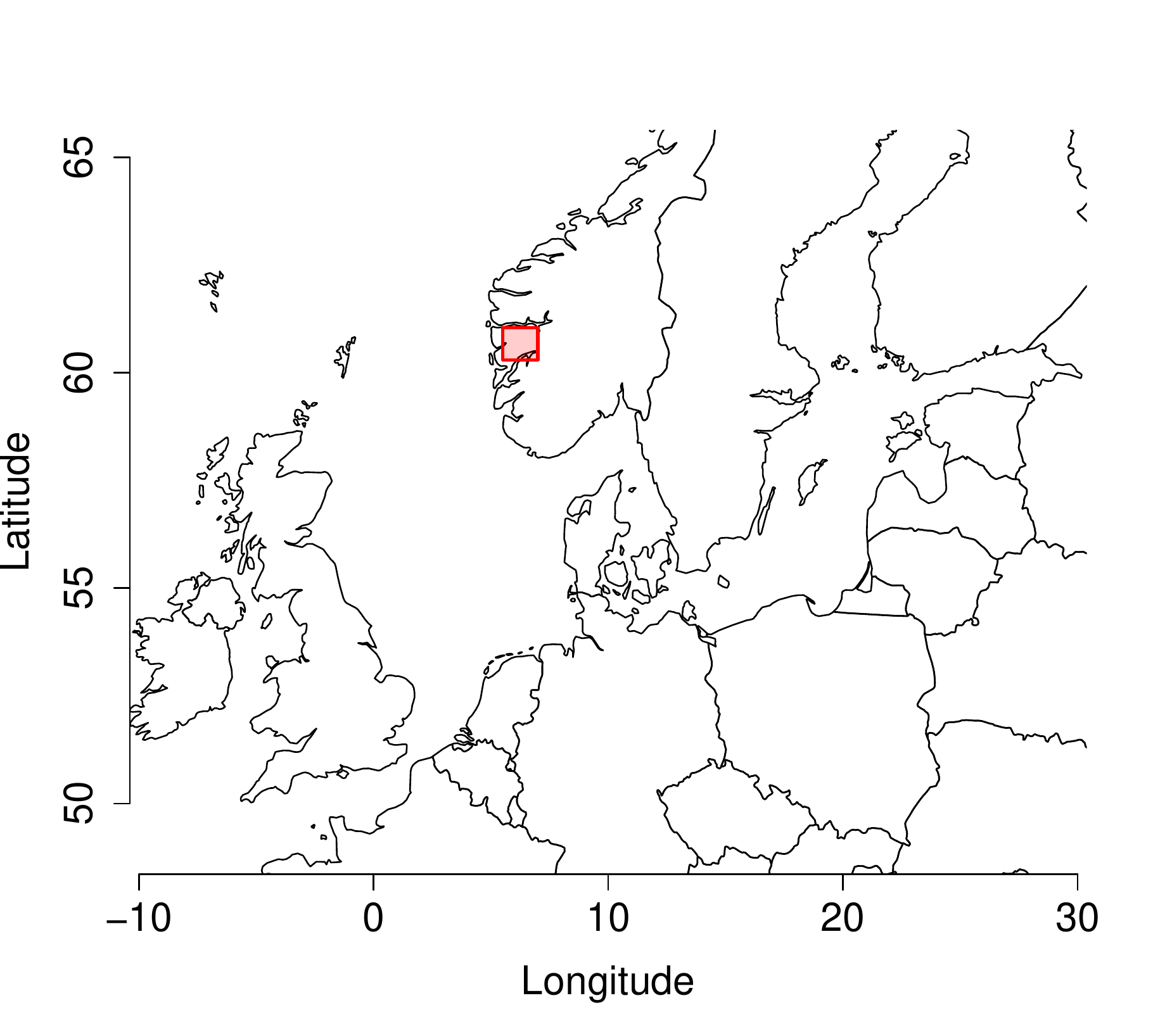}
		\caption{\small The study area is located around Voss in Western Norway.}
		\label{fig:vossieuropa}	
	\end{subfigure}
	~
	\begin{subfigure}[b]{0.45\textwidth}
		\includegraphics[page=1, trim = 0mm 0mm 0mm 0mm, clip, width=8cm]{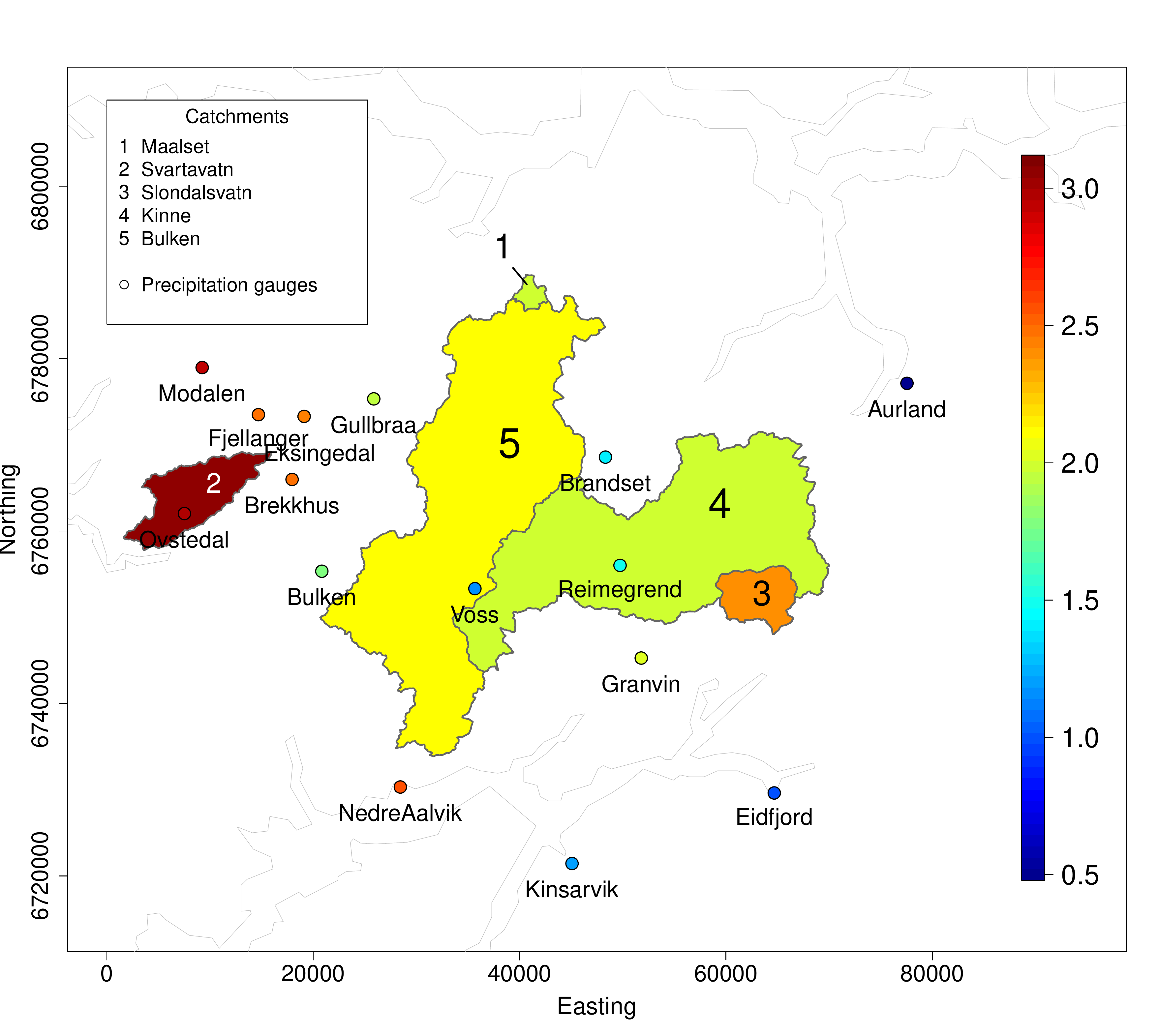}
		\caption{\small Mean annual observations [m/year].}\label{fig:data1}
	\end{subfigure}
	\caption{Mean annual runoff from 5 catchments and mean annual precipitation minus evaporation (m/year) at 15 precipitation gauges for 1988-1997. Catchment 3 is a subcatchment of Catchment 4 and 5, and Catchment 4 is a subcatchment of Catchment 5. Catchment 1 and Catchment 2 don't overlap with any of the other catchments. The coordinate system in Figure \ref{fig:data1} is utm32.}\label{fig:data}
\end{figure*}

When modeling hydrological processes on an annual scale, it is common to use the hydrological definition of a year. The basic water balance equation is given as $P = Q + E + S$, where $P$ is precipitation, $Q$ is runoff, $E$ is evapotranspiration and S is the change in stored water (i.e. snow, or groundwater). A hydrological year is defined such that the storage component in the water balance equation can be neglected, i.e. $S$ is much smaller than $P$ and $Q$.  In Norway a hydrological year starts September 1st and ends August 31st, e.g. 1988 begins September 1st 1987 and ends August 31st 1988. 

In this analysis, we have runoff data from the hydrological years 1988-2014. The dataset was provided by the Norwegian Water Resources and Energy directorate (\textsc{nve}) and consists of annual runoff observations from five catchments where three of them are nested (see Figure \ref{fig:data}). The unit of the data is m/year and gives the spatial average of the runoff within a catchment. The observations from 1988-1997 are used to make statistical inference, while the observations from 1998-2014 are used as a test set for assessing the model's ability to predict runoff for future years.  

The annual runoff data were created by aggregating daily streamflow measurements. The stream gauges that gather the daily observations, don't measure runoff directly: They measure the river's daily stage. Runoff observations are then obtained by using a rating curve that gives the relationship between the stage of the water and the discharge or runoff at a specific point in the stream. The stage-discharge relationship is developed empirically by measuring the discharge across a cross-section of the specific river for a range of stream stages.

Errors in the observed runoff are composed of errors related to the river stage measurement process and errors in the rating curve model. However, on an annual time scale, the river stage measurement errors tend to average out, and the main contribution to errors originates from uncertainties in the rating curve. The dataset provided by \textsc{nve} includes an estimate of the standard deviation of the observation uncertainty for each (annual) runoff observation, and the standard deviations are relatively small ranging from 0.65 \% to 3.2 \% of the corresponding observed value. This information is used to make informative priors for the measurement uncertainties in Section \ref{sec:priors}. We refer to \cite{Reitan2009} for details on how the observation (rating curve model) uncertainty is obtained.

In addition to runoff data, we have precipitation data from 15 precipitation gauges.  Daily precipitation data were downloaded from $\texttt{www.eKlima.no}$ which is a web portal maintained by the Norwegian Meteorological Institute. The observations were aggregated to annual values for the hydrological years 1988-1997. The observed precipitation ranges from 0.55 m/year to 4.6 m/year.

The evaporation data used, originate from the satellite remote sensing-based evapotranspiration algorithm presented in \cite{evap4}. The dataset consists of global monthly land surface evapotranspiration with spatial resolution of 1 degree (longitude, latitude). Evaporation data for the locations of the precipitation gauges around Voss were extracted, and monthly values were aggregated to hydrological years (1988-1997). As the spatial resolution of the gridded evaporation dataset is 1 degree and the study area is rather small, the observed annual evaporation within a specific year is the same for almost all of the precipitation gauges. The observed evaporation ranges from 0.23-0.32 m/year with mean 0.25 m/year and standard deviation 0.02 m/year. This means that approximately $12 \%$ of the annual precipitation evaporates around Voss, which is a small amount in a global perspective. The observations of evaporation must be considered as approximative estimates of the actual evaporation in the area of interest, with large uncertainties.

Figure \ref{fig:data} shows the 5 catchments where we have measurements of runoff and the locations of the 15 precipitation gauges. Mean annual values for areal referenced runoff and point referenced runoff (precipitation-evaporation) for 1988-1997 are included. We see a spatial pattern with high values of annual runoff in the western part of the study area and low values in the eastern part. This pattern is prominent for all years for which we have data, and indicates that climatic spatial effects dominate over annual spatial effects around Voss.

\section{Background}
We propose a Latent Gaussian model (\textsc{lgm}) for annual runoff that is computational feasible due to a stochastic partial differential equation (\textsc{spde}) formulation of Gaussian random fields (\textsc{grf}s). In this section we give a brief introduction of these concepts and other relevant background theory and notation for developing and evaluating the model for annual runoff that is presented in Section \ref{sec:models}.

\subsection{Latent Gaussian Models}\label{sec:LGM}
In this article we suggest a Latent Gaussian model (\textsc{lgm}) for combining point and areal observations of annual runoff. An \textsc{lgm} can be represented in a hierarchical structure consisting of three levels (see e.g. \cite{Gelman}). The first level is the observation likelihood, in this case consisting of two data types $(y_1,...,y_{n})$ and $(z_1,...,z_m)$. The data are observed with conditional independent likelihood $\Pi_{i=1}^{n} \pi(y_i|q_{i},\boldsymbol{\theta_1^y})\Pi_{j=1}^{m} \pi(z_j|Q_{j},\boldsymbol{\theta_1^z})$ given two linear predictors $q_{i}$ and $Q_{j}$, and some parameters ($\boldsymbol{\theta_1^y}$,$\boldsymbol{\theta_1^z}$) which we refer to as hyperparameters.  The two linear predictors depend on the same set of latent variables $\boldsymbol{x}$, but connect the data to the latent field differently, through different projection matrices, e.g. $q_{i}=\boldsymbol{A}_i \boldsymbol{x}$ and $Q_{j}=\boldsymbol{B}_j \boldsymbol{x}$. Here, $\boldsymbol{A}$ and $\boldsymbol{B}$ are matrices that link elements in the latent field to the observations, and $\boldsymbol{A}_i$ and $\boldsymbol{B}_j$ denote row number $i$ and $j$ of the two matrices. The second level of the \textsc{lgm} is formed by the prior of the latent field $\boldsymbol{x}$ and is on the form $ \pi(\boldsymbol{x}|\boldsymbol{\theta_2})\sim \mathcal{N}(\boldsymbol{\mu}(\boldsymbol{\theta_2}),\boldsymbol{\Sigma}(\boldsymbol{\theta_2})),$
i.e. it is Gaussian conditioned on some hyperparameters $\boldsymbol{\theta_2}$. The third level is given by $\pi(\boldsymbol{\theta})$ which is the prior distribution of the hyperparameters  $\boldsymbol{\theta}=(\boldsymbol{\theta_1^y},\boldsymbol{\theta_1^z},\boldsymbol{\theta_2})$.

\subsection{Gaussian random fields}
We use Gaussian random fields (\textsc{grf}s) to model the spatial variability of annual runoff. A continuous field $\{x(\boldsymbol{u});\boldsymbol{u}\in \mathcal{D}\}$ defined on a spatial domain $\mathcal{D}\in \mathcal{R}^2$ is a \textsc{grf} if for any collection of locations $\boldsymbol{u}_1,...,\boldsymbol{u}_n\in \mathcal{D}$ the vector $(x(\boldsymbol{u}_1),...,x(\boldsymbol{u}_n))$ follows a multivariate normal distribution \citep{Cressie}, i.e. $(x(\boldsymbol{u}_1),...,x(\boldsymbol{u}_n)) \sim \mathcal{N}(\boldsymbol{\mu},\boldsymbol{\Sigma})$. The covariance matrix $\boldsymbol{\Sigma }$ defines the dependency structure in the spatial domain, and can be constructed from a covariance function $C(\boldsymbol{u}_i,\boldsymbol{u}_j)$. Furthermore, the dependency structure for a spatial process is often characterized by two parameters: The marginal variance $\sigma^2$ and the range $\rho$. The marginal variance gives information about the spatial variability of the process of interest, while the range gives information about how the correlation between the process at two locations decays with distance. If the range and marginal variance are constant over the spatial domain, we have a stationary \textsc{grf}.

One popular choice of covariance function is the Matérn covariance function which is given by
\begin{equation}\label{eq:Matern}
C(\boldsymbol{u_i},\boldsymbol{u_j})=\frac{\sigma^2}{2^{\nu - 1}\Gamma(\nu)}(\kappa ||\boldsymbol{u_j}-\boldsymbol{u_i}||)^{\nu}K_{\nu}(\kappa||\boldsymbol{u_j}-\boldsymbol{u_i}||),
\end{equation}
where $||\boldsymbol{u_j}-\boldsymbol{u_i}||$ is the Euclidean distance between two locations $\boldsymbol{u_i}, \boldsymbol{u_j} \in \mathcal{R}^d$, $K_{\nu}$ is the modified Bessel function of the second kind and order $\nu>0$, and $\sigma^2$ is the marginal variance \citep{Matern}. The parameter $\kappa$ is the scale parameter, and it can be shown empirically that the spatial range can be expressed as $\rho = \sqrt{8 \nu}/\kappa$, where $\rho$ is defined as the distance where the spatial correlation between two locations has dropped to 0.1 \citep{SPDELindgren}. Using a Matérn \textsc{grf} is convenient because it makes it possible to apply the \textsc{spde} approach to spatial modeling which is outlined in the next subsection.

\subsection{The \textsc{spde} approach to spatial modeling}\label{sec:SPDE}
Making statistical inference and predictions on models including \textsc{grf}s involve matrix operations on the covariance matrix $\boldsymbol{\Sigma}$. This can lead to computational challenges if the covariance matrix is dense. In this paper, we suggest a model for annual runoff that includes not only one, but two \textsc{grf}s. Consequently, some simplifications have to be done to make the model computationally feasible. To achieve this, we use that the exact solution of the \textsc{spde}
\begin{equation}\label{eq:SPDE}
(\kappa^2-\Delta)^{\frac{\alpha}{2}}\tau x(\boldsymbol{u})=\mathcal{W}(\boldsymbol{u}), \quad \boldsymbol{u} \in \mathcal{R}^d, \quad \kappa>0, \quad \nu >0,
\end{equation}
is a Gaussian random field with Matérn covariance function. Here, $\mathcal{W}(\cdot)$ is spatial Gaussian white noise, $\Delta$ is the Laplacian, $\alpha$ is a smoothness parameter, $\kappa$ is the scale parameter in Equation \eqref{eq:Matern}, $d$ is the dimension of the spatial domain and $\tau$ is a parameter controlling the variance. The parameters of the Matérn covariance function in Equation \eqref{eq:Matern} is linked to the \textsc{spde} through
\begin{equation*} \sigma^2=\frac{\Gamma(\nu)}{\Gamma(\alpha)(4\pi)^{d/2}\kappa^{2\nu}\tau^2}; \hspace{20mm} \nu=\alpha - d/2 , 
\end{equation*}
where we will use that $d=2$ and set $\alpha = 2$, such that $\nu$ is fixed to $\nu=1$. The parameter $\nu$ is fixed because it is difficult to identify from data, and $\alpha=2$, $\nu=1$ are commonly used values for these parameters \citep{Rikke1,Chapter6}.

The link between the above \textsc{spde} and the Matérn \textsc{grf}, which was developed by  \cite{Whittle54, Whittle63}, is used by \cite{SPDELindgren} to show that a \textsc{grf} can be approximated by a Gaussian Markov random field (\textsc{gmrf}). This is done by solving the \textsc{spde} in Equation \eqref{eq:SPDE} by the finite element method (\textsc{fem}) (see e.g \cite{FEM}). A \textsc{gmrf} is simply a multivariate Gaussian vector that is parametrized by the precision matrix $\boldsymbol{Q}$, which is the inverse $\boldsymbol{\Sigma}^{-1}$ of the covariance matrix. The term \textsc{gmrf} is most used for Gaussian processes with sparse precision matrices, i.e. matrices that contain many zero elements. The zero elements correspond to Markov properties, in this case conditional independence between locations in the spatial domain. It is convenient to work with \textsc{gmrf}s because there exist  computationally  efficient algorithms for sparse matrix operations \citep{GMRFbook}. Hence, through the \textsc{spde} approach from \cite{SPDELindgren} a \textsc{grf} with a dense precision matrix can be replaced by a \textsc{gmrf} with a sparser precision matrix with computational benefits.

\subsection{\textsc{pc} priors}
As we use a Bayesian approach, the hyperparameters $\boldsymbol{\theta}$ from Section \ref{sec:LGM} must be given prior distributions. For the majority of the hyperparameters we use penalized complexity (\textsc{pc}) priors. \textsc{pc} priors are proper prior distributions developed by \cite{PC1}. The main idea behind \textsc{pc} priors is to penalize the increased complexity induced by deviating from a simple base model. One of the goals is to avoid overfitting.

The \textsc{pc} prior for the precision $\tau$ of a Gaussian effect $\mathcal{N}(0,\tau^{-1})$ has density
\begin{equation}\label{eq:PCprior}
\pi(\tau)=\frac{\lambda}{2}\tau^{-3/2}\exp(-\lambda \tau^{-1/2}),\quad\quad \tau >0, \quad \lambda>0,
\end{equation}
where $\lambda$ is a parameter that determines the penalty of deviating from the base model. The parameter $\lambda$ can be specified through a quantile $u$ and probability $\alpha$ by $\mathrm{Prob}(1/\sqrt{\tau}>u)=\alpha$, where $u>0$, $0<\alpha<1$ and  $\lambda=-ln(\alpha)/u$. Here, $1/\sqrt{\tau}$ is the standard deviation of the Gaussian distribution.

As the range and the marginal variance are easier to interpret than the Matérn covariance function parameters $\kappa$ and $\tau$ in Equation \eqref{eq:Matern}, we parametrize our model through $\rho$ and $\sigma$. For $\rho$ and $\sigma$ we use the prior suggested in \cite{PC2}. This is a joint prior for the spatial range $\rho$ and the marginal variance $\sigma$ constructed from \textsc{pc} priors. The joint prior can be specified through \begin{align*}
\mathrm{Prob}(\rho<u_\rho )=\alpha_\rho; \hspace{20mm} \mathrm{Prob}(\sigma>u_\sigma)=\alpha_\sigma,
\end{align*}
where $u_\rho$, $u_\sigma$, $\alpha_\rho$ and $\alpha_\sigma$ are quantiles and probabilities that must be determined.

\subsection{Evaluating the predictive performance}
To evaluate the predictive performance of the suggested runoff model, we use two criteria: The first criterion is the root mean squared error (RMSE). The RMSE measures the difference between a point prediction $\hat{y}_i$ and the observed value $y_i$ by
\begin{equation*}
\mathrm{RMSE}=\sqrt{\frac{1}{n}\sum_{i=1}^{n}(y_i-\hat{y_i})^2},
\end{equation*}
where $n$ is the total number of pairs of predictions and observations. We use the posterior mean as a point prediction when computing the RMSE. The second criterion is the continuous ranked probability score (CRPS). The CRPS is defined as
\begin{equation*}
\mathrm{CRPS}(F,y)=\int_{-\infty}^{\infty}(F(u)-1\{y\leq u\})^2du,
\end{equation*}
where $F$ is the predictive cumulative distribution and $y$ is the observed value \citep{Gneiting}. The CRPS takes the whole posterior predictive distribution into account, not only the posterior mean or median, and is penalized if the  observed value falls outside the posterior predictive distribution. 
Both the RMSE and the CRPS are negatively oriented, and a smaller value indicates a better prediction.

\subsection{Interpolation by using Top-Kriging}\label{sec:topkriging}
The focus of this article is mainly on highlighting properties of the suggested point and areal runoff model. However, we also compare some of our results to the predictive performance of Top-Kriging. Top-Kriging \citep{topkriging} is one of the leading methods for runoff interpolation. It is a Kriging approach \citep{Cressie} where it is assumed that the variable of interest can be modeled as a \textsc{grf}. A prediction of the target variable at an unobserved location is given by a weighted sum of the available observations, and the interpolation weights are estimated by finding the so-called best linear unbiased estimator (\textsc{blue}). 

In the computation of the interpolation weights, the Top-Kriging approach calculates the covariance between two catchments based on the distance between all the grid nodes in a discretization of the involved catchments. As a consequence, a subcatchment get a higher Kriging weight than a nearby, non-overlapping catchment.  This is different from other Kriging approaches traditionally used in hydrology, for which  streamflow observations have been treated as point referenced (see e.g. \cite{euclid2, euclid3,euclid11}).

While the suggested Bayesian approach for runoff interpolation supports both areal and point observations, Top-Kriging only considers runoff (areal) data. Furthermore, Top-Kriging estimates the covariance (or variogram) empirically, while we take a fully Bayesian approach where the latent field and the parameters are estimated jointly. Another main difference is that Top-Kriging treats each year of runoff data separately, while we can model several years of runoff simultaneously through our two field model.

\section{Statistical Model for Annual Runoff}
\label{sec:models}
In this section we present the proposed \textsc{lgm} for annual runoff which is suitable for combining observations of different spatial support and that has a climatic spatial field that let us quantify long-term spatial variability. 
\subsection{Spatial model for runoff}\label{sec:spatialmodels}
Let the spatial process $\{q_j(\boldsymbol{u}): \boldsymbol{u} \in \mathcal{D} \}$ denote the runoff generating process at a point location $\boldsymbol{u}$ in the spatial domain $\mathcal{D}\in \mathcal{R}^2$ in year $j$. The true runoff generation at
point location $\boldsymbol{u}$ is modeled as
\begin{equation}\label{eq:precip}
q_j(\boldsymbol{u})=\beta_c + c(\boldsymbol{u})+\beta_j + x_j(\boldsymbol{u}),\hspace{3mm} j=1,...,r.
\end{equation}
Here, the parameter $\beta_c$ is an intercept common for all years $j=1,...,r$, while $c(\boldsymbol{u})$ is a spatial effect common for all years. These two model components represent the runoff generation caused by the climate in the study area. Mark that the term climate here covers all long-term effects, i.e. both long-term weather patterns \textit{and} patterns that are repeated due to catchment characteristics. Further, we include a year specific intercept $\beta_j$ and a year specific spatial effect $x_j(\boldsymbol{u})$ for $j=1,..r$ to model the runoff generation due to the annual discrepancy from the climate. Both spatial effects $c(\boldsymbol{u})$ and $x_j(\boldsymbol{u})$ are modeled as \textsc{grf}s with zero mean and Matérn covariance functions given the model parameters; $c(\boldsymbol{u})$ with range parameter $\rho_c$ and marginal variance $\sigma_c^2$, and $x_j(\boldsymbol{u})$ with range $\rho_x$ and marginal variance $\sigma_x^2$. The spatial fields $x_j(\boldsymbol{s})$, j=1,...,r, are assumed to be independent realizations, or replicates of the same underlying \textsc{grf}. The same applies for the year specific intercepts $\beta_j$ which are assumed to be independent and identically distributed as $\mathcal{N}(0,\tau_{\beta}^{-1})$ given the parameter $\tau_{\beta}$ with $\beta_1,...,\beta_r$ being independent realizations of this Gaussian distribution. 

The true mean runoff generated inside a catchment $\mathcal{A}$ in year $j$ can be expressed as
\begin{equation}\label{eq:runoff1}
Q_j(\mathcal{A})=\frac{1}{|\mathcal{A}|}\int_{\boldsymbol{u} \in \mathcal{A}}q_j(\boldsymbol{u})d\boldsymbol{u}, \hspace{3mm} j=1,...,r,
\end{equation}
where $|\mathcal{A}|$ is the area of catchment $\mathcal{A}$. By interpreting catchment runoff as an integral of point referenced runoff $q_j(\boldsymbol{u})$, we obtain a mathematically consistent model where the water balance is conserved for any point in the landscape. 

\subsection{Observation model}
Annual precipitation and evaporation are observed at $n$ locations $\boldsymbol{u}_i \in \mathcal{D}$ for $i=1,..n$ and for $r$ years $j=1,..r$ . The observed annual runoff generation at point location $\boldsymbol{u}_i$, year $j$, is modeled as the difference between the observed annual precipitation $p_{ij}$ and annual evaporation $e_{ij}$,
\begin{equation}\label{eq:pointobs}
y_{ij}=p_{ij}-e_{ij}=q_j(\boldsymbol{u}_i)+\epsilon^y_{ij} \quad i=1,...,n; \quad j=1,...,r,
\end{equation}
where $q_j(\boldsymbol{u}_i)$ is the true annual point runoff from Equation \eqref{eq:precip}. The error terms $\epsilon_{ij}^y$ are independent and identically distributed as $ \mathcal{N}(0,s_{ij}^y \cdot \tau_y^{-1})$ and independent of the other model components. The measurement uncertainties for precipitation and evaporation are assumed to increase with the magnitude of the observed value, and we want to include this assumption in the model. This is done by scaling the precision parameter of the error terms $\tau_{y}$ with a fixed factor $s_{ij}^y$, that is further described in Section \ref{sec:priors}. \\\\
Runoff at catchment level is observed through streamflow data from $K$ catchments denoted $\mathcal{A}_1,...,\mathcal{A}_K$ for $r$ years denoted $j=1,..,r$. We use the following model for the annual runoff observed in catchment $\mathcal{A}_k$ in year $j$

\begin{equation}\label{eq:arealobs}
z_{kj}=Q_j(\mathcal{A}_k)+\epsilon^z_{kj} \quad k=1,...,K; j=1,...,r,
\end{equation}
where $Q_j(\mathcal{A}_k)$ is the true annual areal runoff from Equation \eqref{eq:runoff1}. The measurement errors $\epsilon^z_{kj}$ are independent and identically distributed as $\mathcal{N}(0,s_{kj}^z\cdot \tau_z^{-1})$ and independent of the other model components. As for the point referenced observations, the precision parameter of the error terms $\tau_z$ is scaled with a fixed factor $s_{kj}^z$ that is further described in the next subsection. This way the uncertainty estimates that the data provider \textsc{nve} has for each annual observation can be included in the modeling. 

In Equation \eqref{eq:arealobs} the variable $Q(\mathcal{A})$ defines an areal representation of the annual runoff in catchment $\mathcal{A}_k$. Hence, through the likelihood, the annual runoff in catchment area $\mathcal{A}_k$ is constrained to be close to the actually observed value (with some uncertainty). 

So far we have defined the observation likelihoods for the point and areal observations separately. To construct a joint model for point and areal runoff, we multiply the likelihoods defined in Equation \eqref{eq:pointobs} and \eqref{eq:arealobs} together as described in Section \ref{sec:LGM}. This is done for all $n$ precipitation gauge  locations $i=1,..n$, for all catchments $k=1,..K$ and for all years $j=1,..r$ such that we obtain a model that simultaneously models several years of runoff. Different years are linked together through the climatic part of the model $c(\boldsymbol{u})+\beta_c$ from Equation \eqref{eq:precip}.

\subsection{Prior distributions}\label{sec:priors}
In the suggested model for annual runoff there are 8 parameters ($\tau_y$, $\tau_z$,$\rho_c$, $\rho_x$, $\sigma_c$, $\sigma_x$, $\beta_c$,$\tau_\beta$) that must be given prior distributions. We start by formulating priors for the measurement errors for the point and areal observations.

The variance of the measurement error of the point referenced observation from  precipitation gauge $i$, year $j$, is given by $s_{ij}^y \tau_y^{-1}$ where $\tau_y$ is a hyperparameter and $s_{ij}^y$ is a deterministic value that scales the variance based on expert opinions from \textsc{nve} about the measurement errors for precipitation and evaporation.

The precipitation data are obtained by observing the amount of water or snow that falls into a bucket, but the buckets often fail to catch a large proportion of the actual precipitation, particularly for windy snow events \citep{raingauge, Groisman, Regnusikkerhet}. Based on this and recommendations from \textsc{nve}, the standard deviation of the observation uncertainty for precipitation is assumed to be $10 \%$ of the observed value $p_{ij}$. The evaporation data are obtained from satellite observations and process-models, and are more uncertain than the precipitation data. We assume that the standard deviation for evaporation is $20 \%$ of the observed value $e_{ij}$. The prior knowledge about the point data is used to specify the scale $s_{ij}^y$ for the point observation $y_{ij}$ at location $i$ and year $j$ as follows
\begin{equation*}
s_{ij}^y=\mathrm{Var}(y_{ij})=\mathrm{Var}(p_{ij}-e_{ij})=\mathrm{Var}(p_{ij})+\mathrm{Var}(e_{ij})-2\cdot \mathrm{Cov}(p_{ij},e_{ij})=(0.1 p_{ij})^2+(0.2e_{ij})^2-2\cdot\mathrm{Cov}(p_{ij},e_{ij}).
\end{equation*}
Here, the covariance between the observed precipitation and evaporation is estimated by
\begin{align*}\footnotesize
&\mathrm{Cov}(p_{ij},e_{ij})\hspace{2mm}=\sqrt{\mathrm{Var}(p_{ij})}\cdot \sqrt{\mathrm{Var}(e_{ij})}\cdot \mathrm{Cor}\{ (p_{i1},...,p_{ir}),(e_{i1},...,e_{ir}) \},
\end{align*}
where $Cor\{\cdot,\cdot\}$ is the Pearson correlation between all available observations of precipitation and evaporation at precipitation gauge $i$. Further, we assign the precision $\tau_{y}$ the \textsc{pc} prior from Equation \eqref{eq:PCprior} with $\alpha=0.1$ and $u=1.5$. With this prior, a prior 95 \% credible interval for the standard deviation $\sqrt{s^y_{ij} \tau_y^{-1}}$ of the measurement error for point runoff becomes approximately (0.002-30)\% of the corresponding observed value $y_{ij}$. This interval corresponds well to what \textsc{nve} knows about the measurement uncertainty for precipitation and evaporation.

The same approach is used to make a prior for the variance of the measurement error for the areal referenced observations $z_{kj}$. The precision  $\tau_z$ is given a \textsc{pc} prior with $\alpha=0.1$ and $u=1.5$, while the scale $s_{kj}^{z}$ for catchment $k$, year $j$ is given by
 \begin{equation}\label{eq:scales}
 s_{kj}^z=\mathrm{Var}(z_{kj}).
 \end{equation}
For the streamflow data, information about the variance of the observations is directly available through the dataset provided by \textsc{nve}. These data are inserted into Equation \eqref{eq:scales}.  With the suggested prior, a prior 95 \% credible interval for the standard deviation $\sqrt{s^z_{kj} \tau_z^{-1}}$ of an areal observation, is approximately (0.002,4.0) $\%$ of the corresponding observed value $z_{kj}$. This is an informative prior that just covers the range of values suggested by \textsc{nve}. We have chosen a low prior standard deviation in order to try to put more weight on the runoff observations than to the point observations. There are only 5 areal observations available for each year in the dataset, but 15 point observations, and the aim is to avoid that the more unreliable point data dominate over the areal data. 

For the spatial ranges and the marginal variances of the spatial fields $x_j(\boldsymbol{u})$ and $c(\boldsymbol{u})$, the joint \textsc{pc} prior from \cite{PC2} is used. The \textsc{pc} priors for $\sigma_x$, $\rho_x$, $\sigma_c$ and $\rho_c$ are specified through the following probabilities and quantiles:
\begin{align*}
\mathrm{Prob}(\rho_x<10 \text{ km})=0.1, \hspace{2mm} \mathrm{Prob}(\sigma_x>2 \text{ m/year})=0.1,\\ \mathrm{Prob}(\rho_c<10 \text{ km} )=0.1,\hspace{2mm} \mathrm{Prob}(\sigma_c>2 \text{ m/year})=0.1.
\end{align*} 

The percentages and quantiles are chosen based on expert knowledge about the spatial variability in the area. The study area is approximately 80 km $\times$ 80 km, and it is reasonable to assume that there is a correlation larger than 0.1 between two locations that are less than 10 km apart. Furthermore, the spatial variability in the study area is large, and we can observe runoff values from 0.8 m/year to 3.2 m/year within the same year. However, it is reasonable to assume that the marginal standard deviation of the runoff generating process does not exceed 2 m/year. The parameters of the climatic \textsc{grf} $c(\boldsymbol{u})$ and the annual \textsc{grf} $x_j(\boldsymbol{u})$ are given the same prior as it is difficult to identify if the spatial variability mainly comes from climatic processes or from annual variations. We also want the data to determine which of the two effects that dominates in the study area.

As described in Section \ref{sec:spatialmodels}, the year specific intercept $\beta_j$ has prior $\mathcal{N}(0 , \tau_{\beta}^{-1})$ for all years $j=1,..r$. Its precision $\tau_{\beta}$ is given the \textsc{pc} prior from Equation \eqref{eq:PCprior} with $u=10$ and $\alpha=0.2$. This is a weakly informative wide prior with a prior $95 \%$ interval (0.002,40.5) m/year for the standard deviation $\sqrt{\tau_{\beta}^{-1}}$ of $\beta_j$. Finally, the climatic intercept $\beta_c$ is given a normal prior,  $\beta_c\sim \mathcal{N}(2,0.5^2)$. This gives a prior $95 \%$ credible interval of (1.0,3.0) m/year for $\beta_c$ which covers all reasonable mean values of annual runoff around Voss. 

\subsection{Inference}
\begin{figure*}
	\centering
	\begin{subfigure}[b]{0.4\textwidth}
		\includegraphics[page=1, trim = 20mm 20mm 10mm 10mm, clip, width=4cm]{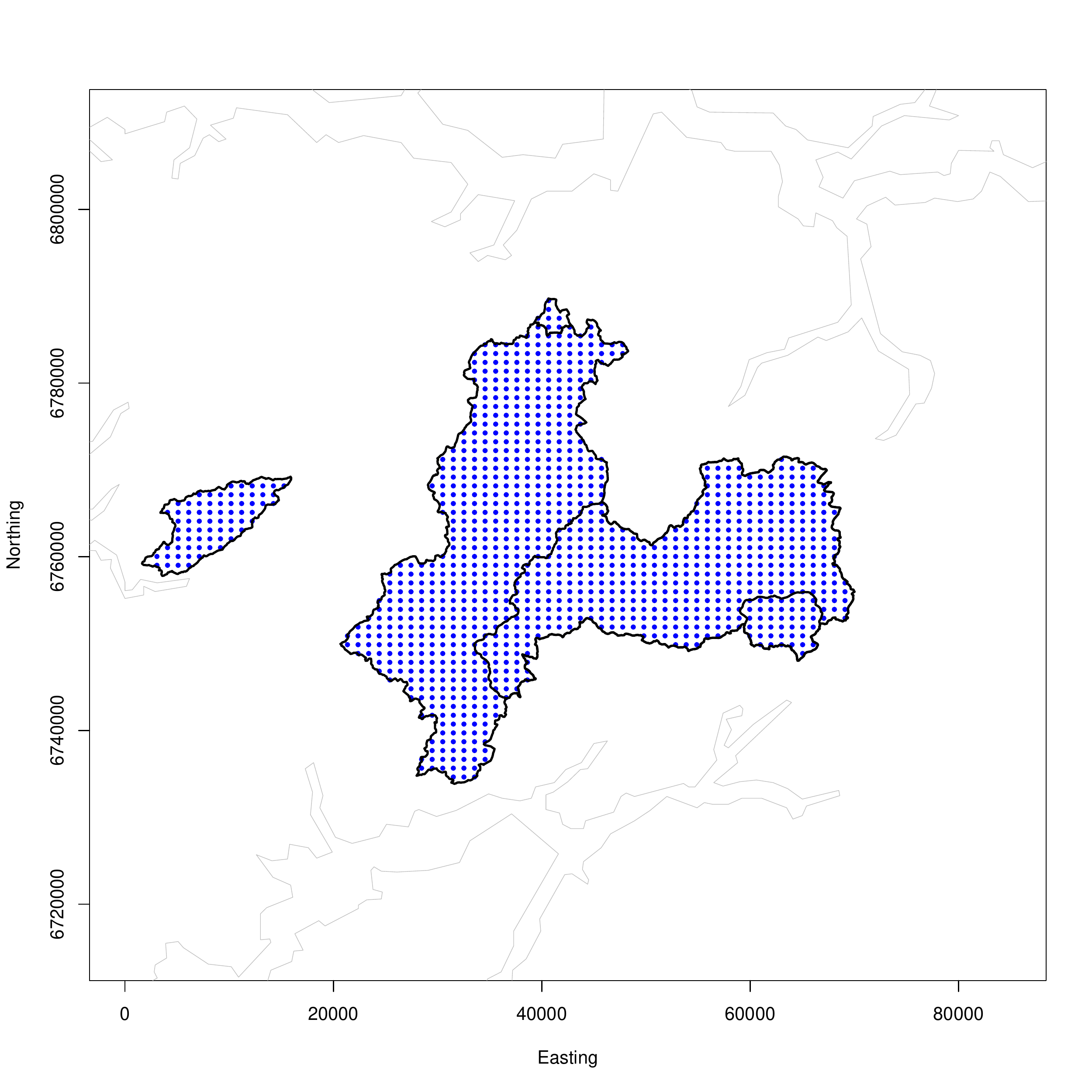}
		\caption{\small Discretization of catchments used to model areal runoff. We use a regular grid with 1 km spacing.}
		\label{fig:catchgrid}
	\end{subfigure}
	~ 
	\begin{subfigure}[b]{0.5\textwidth}
		\includegraphics[page=1, trim = 20mm 10mm 1mm 20mm, clip, width=6cm]{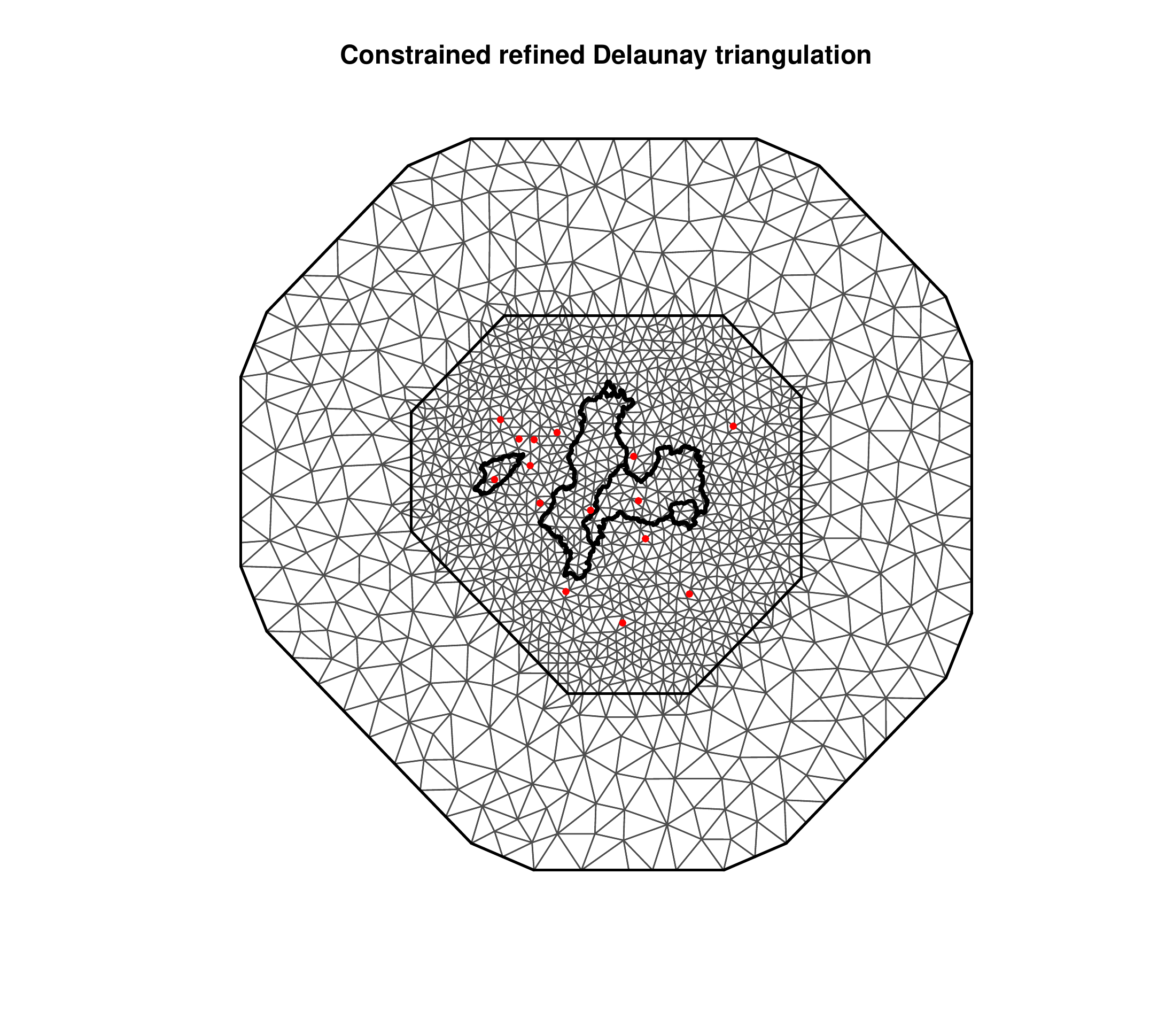}
		\caption{\small Triangulation of the spatial domain used for FEM. The red points are the locations of the precipitation gauges.}
		\label{fig:catchmesh}
	\end{subfigure}
	\caption{Discretization and triangular mesh used to make the model computationally feasible.}
\end{figure*}

In order to make the model computationally feasible, some simplifications of the suggested model are necessary. In Section \ref{sec:spatialmodels} the annual runoff for a catchment $\mathcal{A}_k$ was modeled as the integral of point referenced runoff over the catchment area. In practice, the integral in Equation \eqref{eq:runoff1} is calculated by a finite sum over a discretization of the target catchment. More specifically, let $\mathcal{L}_k$ denote the discretization of catchment $\mathcal{A}_k$. The total annual runoff in catchment $\mathcal{A}_k$ in year $j$ is approximated by 
\begin{align}
Q_j(\mathcal{A}_k)&=\frac{1}{N_k}\sum_{\boldsymbol{u} \in \mathcal{L}_k}q_j(\boldsymbol{u}),
\label{eq:runoffbasis}
\end{align}
where $N_k$ is the total number of grid nodes in $\mathcal{L}_k$ and $q_j(\boldsymbol{u})$ is the point runoff at grid node $\boldsymbol{u} \in \mathcal{L}_k$. It is important that a subcatchment shares grid nodes with the main catchment in order to preserve the water balance. The discretization used in this analysis has 1 km spacing and is shown in Figure \ref{fig:catchgrid}. 

The model suggested for annual runoff, is a latent Gaussian model with the structure described in Section \ref{sec:LGM}. Modeling annual runoff as a \textsc{lgm} is convenient because it allows us to use integrated nested Laplace approximations (\textsc{inla}) to make inference and predictions. \textsc{inla} can be used for making Bayesian inference on \textsc{lgm}s and is a faster alternative to \textsc{mcmc} algorithms \citep{mcmc}. The approach is based on approximating the marginal distributions by using Laplace or other analytic approximations, and on numerical integration schemes. The main computational tool is the sparse matrix calculations described in \cite{GMRFbook}, such that in order to work fast, the latent field  of the \textsc{lgm} should be a \textsc{gmrf} with a sparse precision matrix. In our case, sparsity is obtained by using the \textsc{spde} approach from Section \ref{sec:SPDE} to approximate the \textsc{grf}s $x_j(\boldsymbol{u})$ and $c(\boldsymbol{u})$ by \textsc{gmrf}s. This is done through the finite element method (\textsc{fem}), and the triangulation used for \textsc{fem} is shown in Figure \ref{fig:catchmesh}. In order to obtain accurate approximations of the underlying two \textsc{grf}s, this triangular mesh must be dense enough to capture the rapid spatial variability of annual runoff around Voss. If the mesh is too coarse, unrealistic results such as negative runoff can occur, or we can get into numerical problems.

The R-package \texttt{r-inla} was used to make inference and predictions for the suggested model. This package provides a user-friendly interface for applying \textsc{inla} and the \textsc{spde} approach to spatial modeling without requiring that the user has deep knowledge about \textsc{spde}s. See \texttt{r-inla.org} or \cite{Chapter6} and \cite{Elias} for tutorials and examples. In particular, \cite{areadata1} is recommended for a description of how a model with point and areal data can be implemented in \texttt{r-inla}. 

\section{Case study of annual runoff in Voss}\label{sec:casestudy}
The model presented in Section \ref{sec:models} is used to explore the predictability of annual runoff in the Voss area. Recall that the main goals are to investigate how the predictions are affected by the two different observation types (point and areal data), to demonstrate how the water balance considerations can be beneficial, and to explore the properties of the climatic part of the model. To address this, we perform four tests that are inspired by common applications in hydrology. These are presented in the next subsection. In Section \ref{sec:results} the results from the tests are presented and discussed.

\subsection{Model evaluation}\label{sec:eval}
To explore how the two different observation types influence the predictions of annual runoff around Voss, we compare three observation designs: An observation design where only point referenced observations are included in the likelihood ($P$), an observation design where only areal referenced observations are included in the likelihood ($A$) and an observation design where all available observations are included in the likelihood ($P+A$). Recall that using only areal observations ($A$) corresponds to what typically has been done in hydrological applications, and we want to investigate if we can improve the predictability of runoff by also including point observations in the likelihood ($P+A$). Including $P$ as an observation design gives information about what influence the point data have on the predictions. The three observation designs are evaluated according to four tests that are described as follows:\\

\textbf{T1 - Inference:} The model from Section \ref{sec:models} is fitted to all available observations between 1988 and 1997 from Figure \ref{fig:data}. This is done for $P$, $A$ and $P+A$, such that we get information about how the different observation types affect the posterior estimates of the parameters.\\

\textbf{T2 -Spatial predictions in ungauged catchments:} In hydrological applications, the main interest is on estimating runoff at catchment level. Motivated by this, we perform spatial predictions of annual runoff for each of the five catchments  $\mathcal{A}_1,...,\mathcal{A}_5$ by leave-one-out-cross-validation for $P$, $A$ and $P+A$. That is, data from the target catchment are left out and the catchment of interest is treated as ungauged. Runoff predictions are done for the target catchment for 1988-1997 and are based on observations from the remaining 4 catchments and/or point data from 1988-1997. The predictive performance is assessed by computing the RMSE and CRPS for each catchment based on the 10 years of predictions. 

In \textbf{T2}, we also compare our results to the Top-Kriging approach described in Section \ref{sec:topkriging}. For Top-Kriging, we fit the default covariance function (or variogram) from the R package \texttt{rtop}. This is a multiplication of a modified exponential and fractal variogram model \citep{topkriging}. Recall that Top-Kriging only supports areal referenced (runoff) observations.\\

\textbf{T3u- Future predictions in ungauged catchments:} In \textbf{T2} we estimate the runoff that was generated in ungauged catchments in the past. However, quantifying the annual runoff we can expect in the future is more interesting for most hydrological applications. In \textbf{T3u} we therefore estimate annual runoff for a future year, i.e. for a year for which there are no observations of runoff, precipitation or evaporation. For an unobserved year ($j>10$) the posterior means of the year specific effects $\beta_j$ and $x_j(\boldsymbol{u})$ are zero. Thus, the posterior predicted future runoff is given by the posterior means of the climatic components $\beta_c$ and $c(\boldsymbol{u})$. However, all four model components as well as the observation uncertainty contribute to the predictive uncertainty.

In \textbf{T3u} the catchment of interest is treated as ungauged and left out of the dataset, and we use the remaining observations from 1988-1997 to predict annual runoff for 1998-2014. This is done for catchment $\mathcal{A}_1,...,\mathcal{A}_5$ in turn. The predictive performance is evaluated by computing the RMSE and CRPS for predictions of runoff for each of the 5 catchments for 17 future years. The average RMSE and CRPS over the 5 catchments are used as summary scores. As the posterior mean for an unobserved year is given by the posterior mean of the climatic effects  $\beta_c$ and $c(\boldsymbol{u})$, this test lets us quantify the climatology in the study area.\\

\textbf{T3g - Future predictions in partially gauged catchments:} We predict annual runoff in catchment $\mathcal{A}_1,...,\mathcal{A}_5$ for a future year as in \textbf{T3u}. However, we allow the observation likelihood to contain 1 to 10 annual runoff observations from the catchment in which we want to predict runoff. This way, we assess the model's ability to exploit short records of runoff, which is a property enabled by the climatic component of the model. We denote this test \textbf{T3g}, for gauged, as opposed to \textbf{T3u} for ungauged.

The test is carried out by drawing $i$ observations between 1988-1997 randomly from the target catchment. Next, these observations are used together with the other point and/or areal observations of $P$, $A$ and $P+A$ from 1988-1997 to predict the annual runoff in 1998-2014 for this particular catchment. As the experimental results might depend on which runoff observations we pick from the target catchment, the experiment is repeated 10 times such that different observations are included for each experiment.

The above procedure is carried out with an increasing number of years included in the short record, i.e. for $i\in \{1,2,3,4,5,6,7,8,9,10\}$. The predictive performance is then  evaluated for each $i$ by computing the RMSE and CRPS for each catchment  $\mathcal{A}_1,...,\mathcal{A}_5$ based on 17 years of future predictions. The average RMSE and CRPS over 5 catchments and 10 experiments are reported as summary scores. \\

For our experiments we use the posterior mean as the predicted value when computing the RMSE. Furthermore, when evaluating the CRPS and when computing the coverage of the predictions, we assume that the posterior distributions are Gaussian with mean given by the posterior mean and standard deviation given by the posterior standard deviation. In the posterior standard deviation, we take the measurement uncertainty given by $s_{kj}^z\tau_z^{-1}$ into account, in addition to the uncertainty of the model components of the linear predictor in Equation \eqref{eq:precip}. The Gaussian distribution should be a good approximation for the resulting posterior distributions as they typically are symmetric with neither particularly short or long tails.

\subsection{Results from the case study} \label{sec:results}
 \begin{figure*}	[h!!]
 	
 	\begin{subfigure}[b]{1\textwidth}
 		\includegraphics[width=15cm]{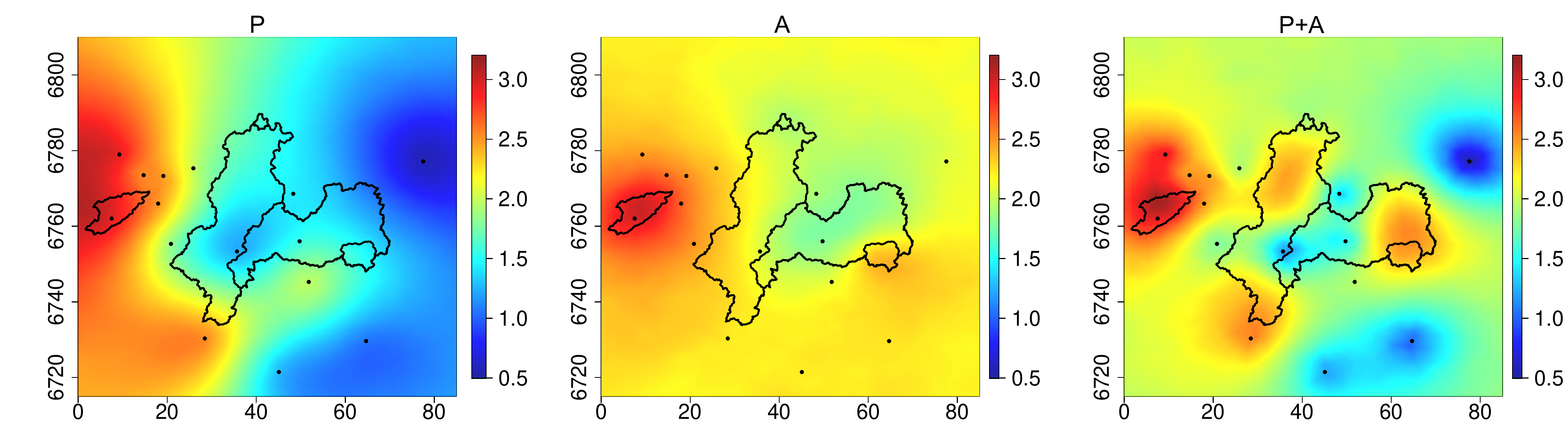}
 		\caption{\small Posterior mean [m/year].}
 		\label{fig:future_eta_mean}
 	\end{subfigure}\\
 	
 	
 	\begin{subfigure}[b]{1\textwidth}
 		\includegraphics[width=15cm]{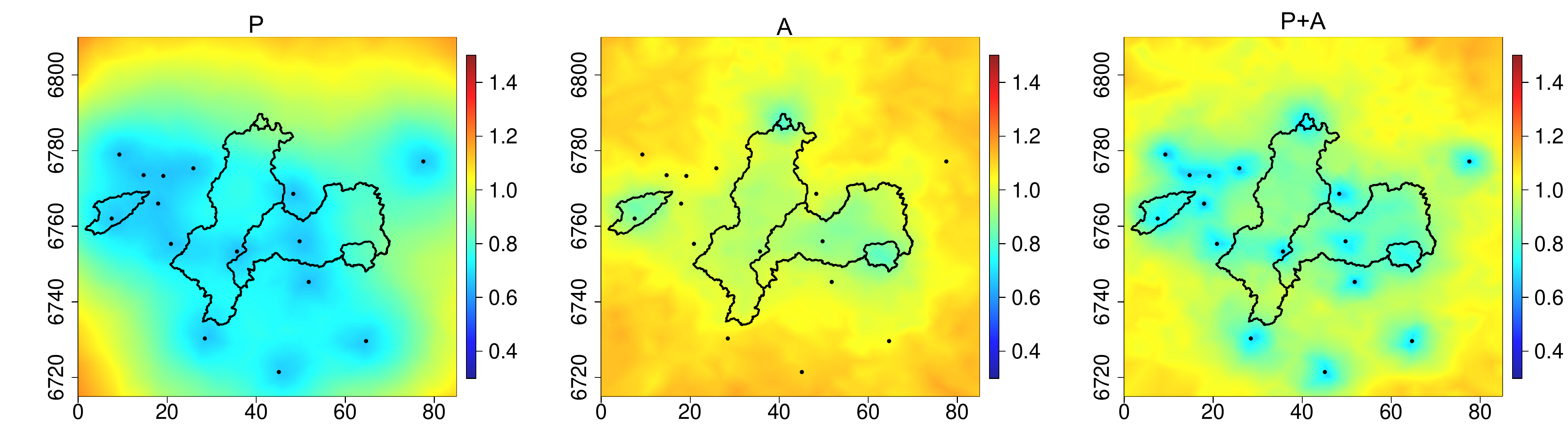}
 		\caption{\small Posterior standard deviation [m/year].}
 		\label{fig:future_eta_sd}
 	\end{subfigure}
 	~
 	\caption{\small Posterior mean and standard deviation for annual runoff for a future unobserved year when all available observations of point observations ($P$, left), areal ($A$, middle) and both point and areal observations ($P+A$, right) from 1988-1997 are used, i.e. all catchments are treated as gauged for $A$ and $P+A$, but ungauged for $P$ (test \textbf{T1}).}\label{fig:futurecompare}
 \end{figure*}
 
 We now present the results from the case study for our four tests \textbf{T1}, \textbf{T2}, \textbf{T3u} and \textbf{T3g} in turn. 
 
 Table \ref{tab:parameter_estimates} shows the posterior medians and the 0.025 and 0.975 quantiles for the hyperparameters for $P$ (point observations), $A$ (areal observations) and $P+A$ (point and areal observations) when all respective available observations from 1988-1997 are used to make inference (\textbf{T1}). In general, $P$ gives lower runoff values with a posterior median of the climatic intercept $\beta_c$ equal to 1.87 m/year compared to $A$ giving $\beta_c$ equal to $2.21$ m/year. Furthermore, the posterior median of the marginal standard deviation of the climatic \textsc{grf} $\sigma_c$ is  considerably larger for $P$ with $\sigma_c=0.97$ m/year compared to $A$ and $P+A$ which give posterior medians 0.63 m/year and 0.76 m/year respectively. The posterior median of the range of the climatic \textsc{grf} $\rho_c$ is also larger for $P$ with 70 km compared to values around 20 km for $A$ and $P+A$. 
 
 The spatial runoff patterns corresponding to these parameter values are shown in Figure \ref{fig:futurecompare}. These figures show the posterior mean and standard deviation for runoff for an unobserved, future year. We see that larger values for $\rho_c$ and $\sigma_c$ lead to a more prominent spatial pattern for $P$ with large runoff values in the western part of the study area and lower values in the eastern part. A high climatic range $\rho_c$ also leads to a reduction of the posterior predictive uncertainty in a larger part of the study area for $P$, as can be seen in Figure \ref{fig:future_eta_sd}.  The maps show that the choice of observation scheme ($P$, $A$ or $P+A$) has a large impact on the resulting predictions of annual runoff in terms of posterior mean and/or posterior standard deviation. 
 
 \begin{table}\small	\centering
 		\caption{\small Posterior median (0.025 quantile, 0.975 quantile) when all available point (P), areal (A) and both point and areal (P+A) referenced observations from 1988-1997 are used for making inference (test \textbf{T1}). The precision parameters are transformed to standard deviations to make them more interpretable. Recall that the posterior estimates of the standard deviations $1/\sqrt{\tau_y}$ and $1/\sqrt{\tau_z}$ of the measurement uncertainties are multiplied with the scales from Section \ref{sec:priors} in order to obtain the final posterior observation uncertainty with unit [m/year].\vspace{3mm}}\label{tab:parameter_estimates}
 	\begin{tabular}{llllllllll}  
 		\toprule
 		Parameter [unit]&& \multicolumn{3}{c}{Posterior median (0.025 quantile, 0.975 quantile)}\\
 		\cmidrule(r){1-1} 	\cmidrule(r){3-5} 	
 		&	&P    & A & P+A \\
 		\midrule
 		$\rho_x$ [km]& &	 236 (148, 379) & 104 (32, 262)  &102 (41, 249)\\
 		$\sigma_x$ [m/year]&&	0.27 (0.20, 0.38) &0.34 (0.18, 0.56)&0.29 (0.19, 0.44)\\
 		$\rho_c$ [km]&	&70 (30, 180)      & 25 (9, 74)  & 20 (9, 46)  \\
 		$\sigma_c$ [m/year]&	&0.97 (0.56, 1.79)      & 0.63 (0.34, 1.34) & 0.76 (0.53, 1.1)\\
 		$\beta_c$ [m/year]&&1.87 (1.13, 2.68)   & 2.21 (1.57, 2.82)   &1.96 (1.40, 2.50)\\
 		$1/\sqrt{\tau_y}$ [unitless]&	& 0.48 (0.40, 0.57)     & $\times$   & 0.37 (0.22, 0.54)     \\
 		$1/\sqrt{\tau_z}$ [unitless]&	&$\times$       &3.6 (2.3, 5.1)  & 5.3 (3.8,6.8) \\
 		$1/\sqrt{\tau_\beta}$ [$\text{m}/\text{year}$]&	&0.26 (0.01, 0.78)  & 0.61 (0.31,1.0)   & 0.48 (0.24, 0.75)     \\
 		\bottomrule
 	\end{tabular}
 \end{table}

\begin{figure*}[h!]		\centering
	\begin{subfigure}[b]{0.5\textwidth}
		\includegraphics[page=1, trim = 0mm 0mm 0mm 0mm, clip, width=7cm]{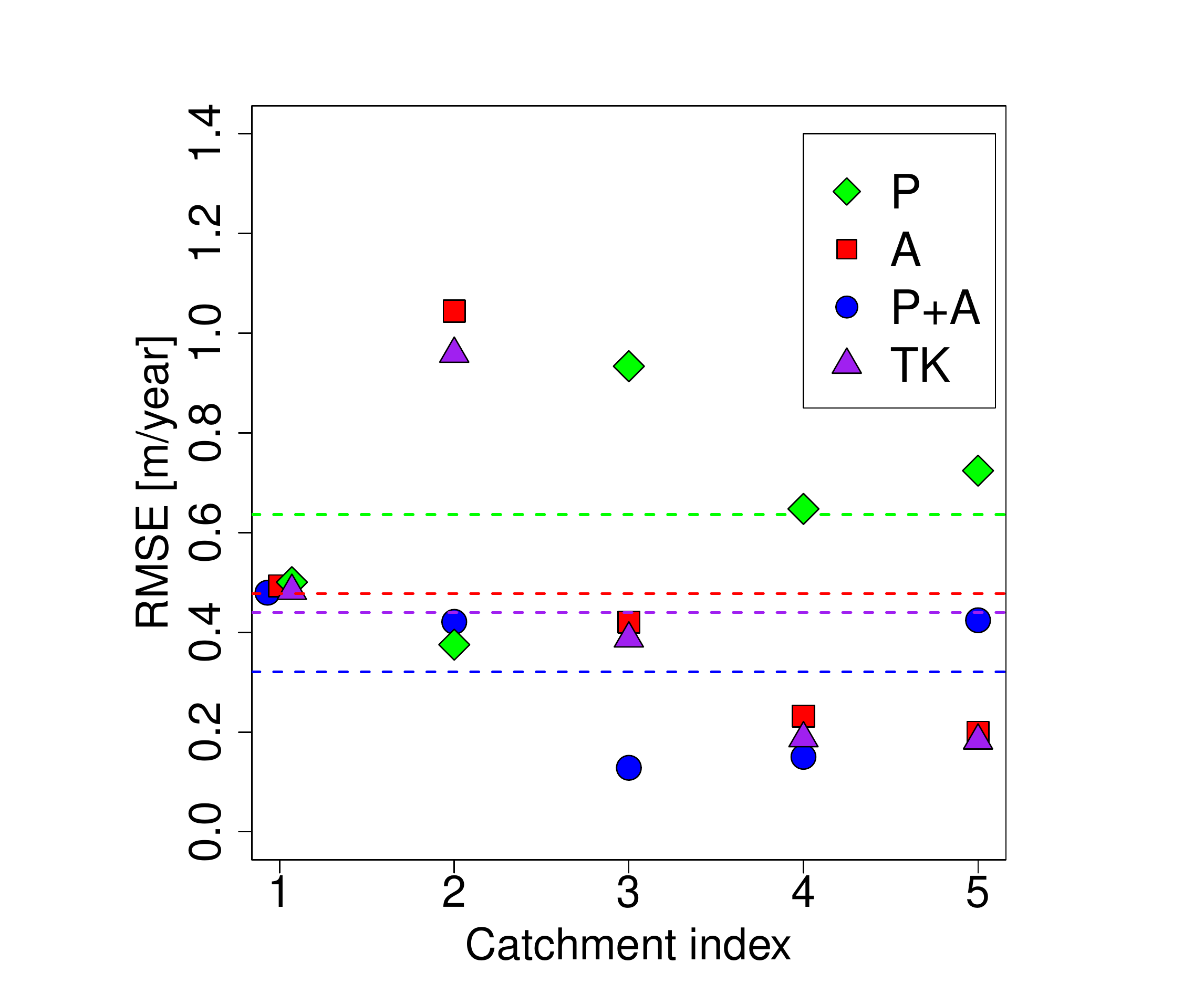}
		\caption{\small RMSE.}
		\label{fig:RMSE_real}
	\end{subfigure}
	\begin{subfigure}[b]{0.4\textwidth}
		\includegraphics[page=1, trim = 0mm 0mm 0mm 0mm, clip, width=7cm]{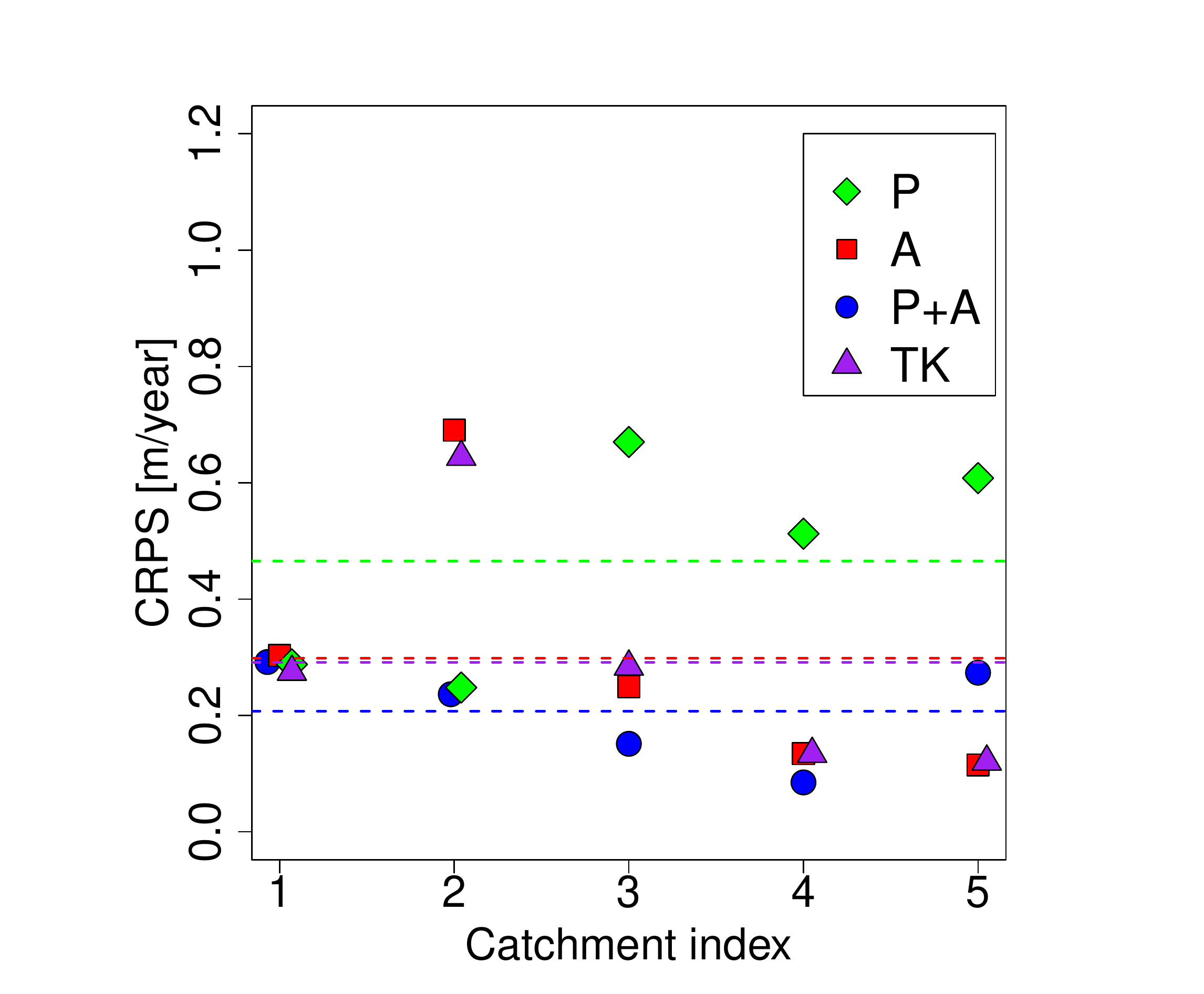}
		\caption{\small CRPS.}
		\label{fig:CRPS_real}
	\end{subfigure}
	~
	\caption{\small Predictive performance for spatial predictions of runoff in 1988-1997 when the target catchment is treated as ungauged (test \textbf{T2}) for P, A and P+A. Here, we have also included results from the reference method Top-Kriging (TK) that only considers areal observations. Dashed lines mark the average performance over all catchments.}
	\label{fig:RMSE_CRPS_real}
\end{figure*}

In \textbf{T2} we perform spatial predictions of annual runoff in 1988-1997 for a catchment that is left out of the dataset. The predictive performance for spatial predictions is summarized in Figure \ref{fig:RMSE_CRPS_real}. For four out of five catchments, $P+A$ gives the lowest RMSE and CRPS, or a RMSE and CRPS that is approximately as for $A$, $P$ or Top-Kriging (TK). We see that the Top-Kriging approach performs similar to $A$, which is reasonable as Top-Kriging only considers areal observations and uses a similar interpretation of covariance as our suggested model.

\begin{figure*}		\centering
		\hspace{17mm} $P$\hspace{39mm}  $A$ \hspace{34mm}  $P+A$\\ \vspace{-3mm}
		
		\includegraphics[page=1,trim = 0mm 1mm 2mm 11mm,clip, width=15cm]{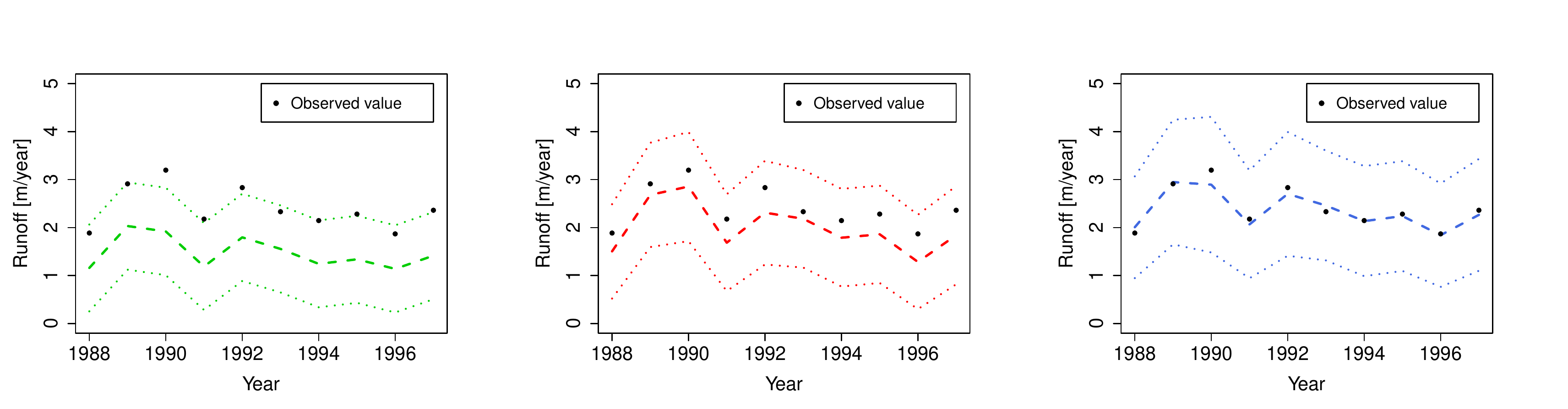}

	\caption{ \small The posterior mean for spatial predictions in Catchment 3 (test \textbf{T2}) with corresponding 95 $\%$ posterior prediction intervals for observation design $P$ (left), $A$ (middle) and $P+A$ (right).}\label{fig:Catchment3_T2}
\end{figure*}

In Figure \ref{fig:RMSE_CRPS_real} we particularly highlight Catchment 3 because it provides an example of how the water balance properties of the model can be beneficial. Figure \ref{fig:RMSE_CRPS_real} shows that for Catchment 3, $P$ gives a RMSE around 0.9, while $A$ gives a RMSE around 0.4. Considering the posterior prediction intervals for Catchment 3 in Figure \ref{fig:Catchment3_T2}, we see that $P$ leads to an underestimation of the annual runoff. This can be explained by looking at the observations in Figure \ref{fig:data}: The point observations close to Catchment 3 all have mean values lower than the true mean annual runoff in this catchment. Next, considering the results for the areal observations (A), Figure  \ref{fig:Catchment3_T2} shows that also these lead to an underestimation of Catchment 3's runoff. Intuitively, we would thus expect that combining $P$ and $A$ would result in underestimation. Instead, we get a large improvement in the predictions in Figure \ref{fig:Catchment3_T2} when $P$ and $A$ are combined, with a RMSE around 0.1 (Figure \ref{fig:RMSE_CRPS_real}). The predictions for Catchment 3 also turn out to be larger than any of the nearby observed values.

The result can be understood by looking at the nested structure of the catchments in the dataset. Catchment 4 and Catchment 5 cover Catchment 3, and through our model formulation they put constraints on the total runoff in this area. As Figure \ref{fig:data} shows, there are two precipitation gauges inside Catchment 5 for which the point runoff generated is lower than the mean annual runoff in the surrounding two catchments. To preserve the water balance, the predicted annual runoff in the remaining parts of Catchment 4 and Catchment 5 has to be larger than any of the values that are observed in the surrounding area. This interaction between nested areal observations and point observations makes the model able to correctly identify Catchment 3 as a wetter catchment than any of the nearby catchments, and we have demonstrated that we have a geostatistical model that does more than smoothing.

 This does not mean that the interaction between point and areal observations always lead to improved predictions (see e.g. Catchment 5 in Figure \ref{fig:RMSE_CRPS_real}). However, overall the results in Figure \ref{fig:RMSE_CRPS_real} show that on average we benefit from including all available data ($P+A$) in the analysis when making spatial predictions, and that using only point observations gives poor predictions. $P$ performs considerably worse than $A$, $P+A$ and Top-Kriging for three of the catchments (Catchment 3, 4 and 5)

\begin{figure*}
	\centering	
	\includegraphics[page=1, trim = 0mm 1mm 1mm 0mm, clip, width=14cm]{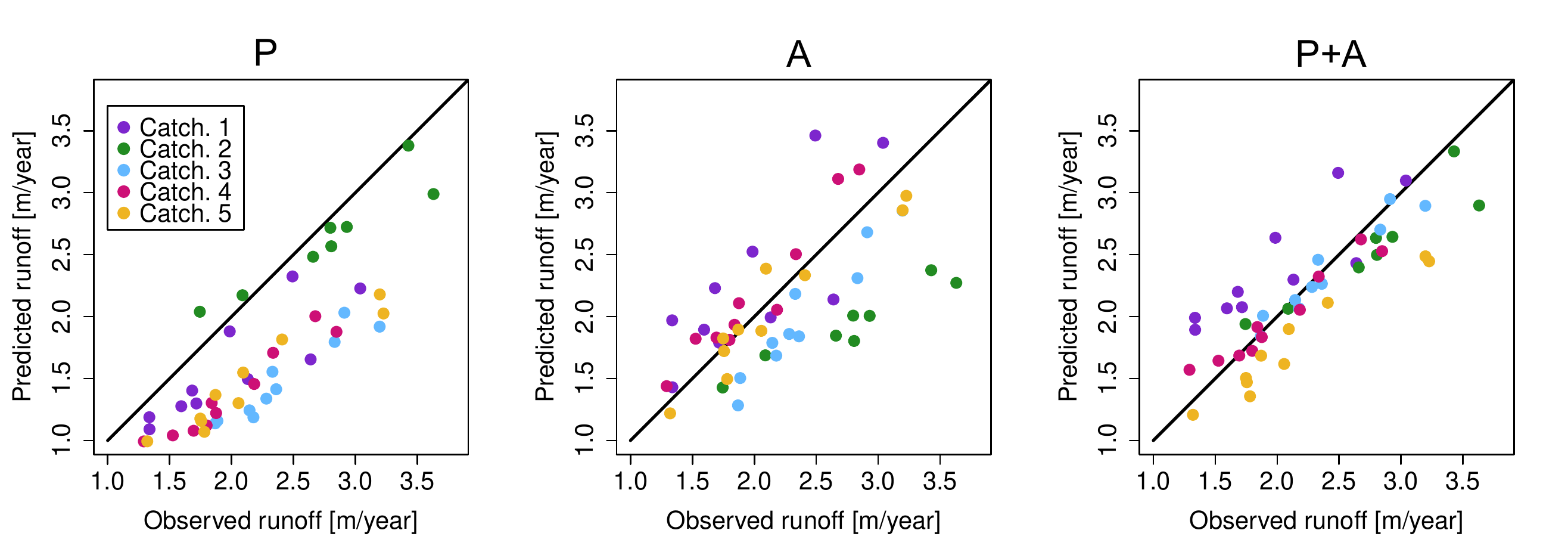}
	\caption{\small The posterior mean for spatial predictions in ungauged catchments for $P$ (left), $A$ (middle) and $P+A$ (right) compared to the corresponding observed value (\textbf{T2}).}\label{fig:scatter_real}
\end{figure*}

The scatterplots in Figure \ref{fig:scatter_real} compare the spatial predictions from 1988-1997 (\textbf{T2}) to the actual observations for each (ungauged) catchment for $P$, $A$ and $P+A$. Overall, observation designs $A$ and $P+A$ provide predictions that are symmetric around the corresponding observed runoff. However, if we look more closely at the predictions for each catchment, we see that $A$ and $P+A$ tend to either overestimate or underestimate the annual runoff within a catchment. This is seen most clearly for Catchment 1 where the annual runoff is overestimated for $A$ and $P+A$, and for Catchment 2 where the runoff is underestimated for $A$. Top-Kriging is not visualized here, but this reference approach gives similar results as observation scheme $A$. 

The results in Figure \ref{fig:scatter_real} show that the same systematic prediction error typically is done each year for a specific catchment. The biases are however small enough that the actual observations are covered by the corresponding $95 \%$ posterior prediction intervals for $A$ and $P+A$ for most catchments. This can be seen in Table \ref{tab:coverage_spatPred}.  %

For $P$ the situation is different: Figure \ref{fig:scatter_real} shows that the annual runoff is underestimated for all catchments. In addition, the posterior standard deviation for runoff is typically unrealistically small for $P$ contributing to narrow posterior prediction intervals. Large biases combined with small posterior standard deviations lead to a low empirical coverage for the spatial predictions for $P$, and on average the coverage of a $95 \%$ posterior prediction interval is as low as $42 \%$. For $P$, neither the posterior mean nor the posterior variance reflects the properties of the underlying process. 
\begin{table}\caption{\small The proportion of the observations that falls into the corresponding $95 \%$ posterior prediction interval for spatial predictions of runoff (\textbf{T2}) in catchment $\mathcal{A}_1,..,\mathcal{A}_5$ for 1988-1997 when the target catchment is treated as ungauged.}	\label{tab:coverage_spatPred}
	
	\small
		\centering
	\begin{tabular}{lllllll}
		\hline
		& $\mathcal{A}_1$ & $\mathcal{A}_2$   & $\mathcal{A}_3$  & $\mathcal{A}_4$   & $\mathcal{A}_5$   & All \\
		\hline
		$P$     & 1 & 0.5 & 0.5 & 0.1 & 0   & 0.42    \\
		$A$     & 1 & 0.7 & 1   & 0.9   & 1   & 0.92    \\
		$P+A$ & 1 & 1 & 1   & 1   & 0.60 & 0.92  \\
		\hline  
	\end{tabular}
\end{table}

\begin{figure*}	[h!!!]
	\centering 
	\begin{subfigure}[b]{0.45\textwidth}
		\includegraphics[page=1, trim = 5mm 2mm 0mm 1mm, clip, width=6cm]{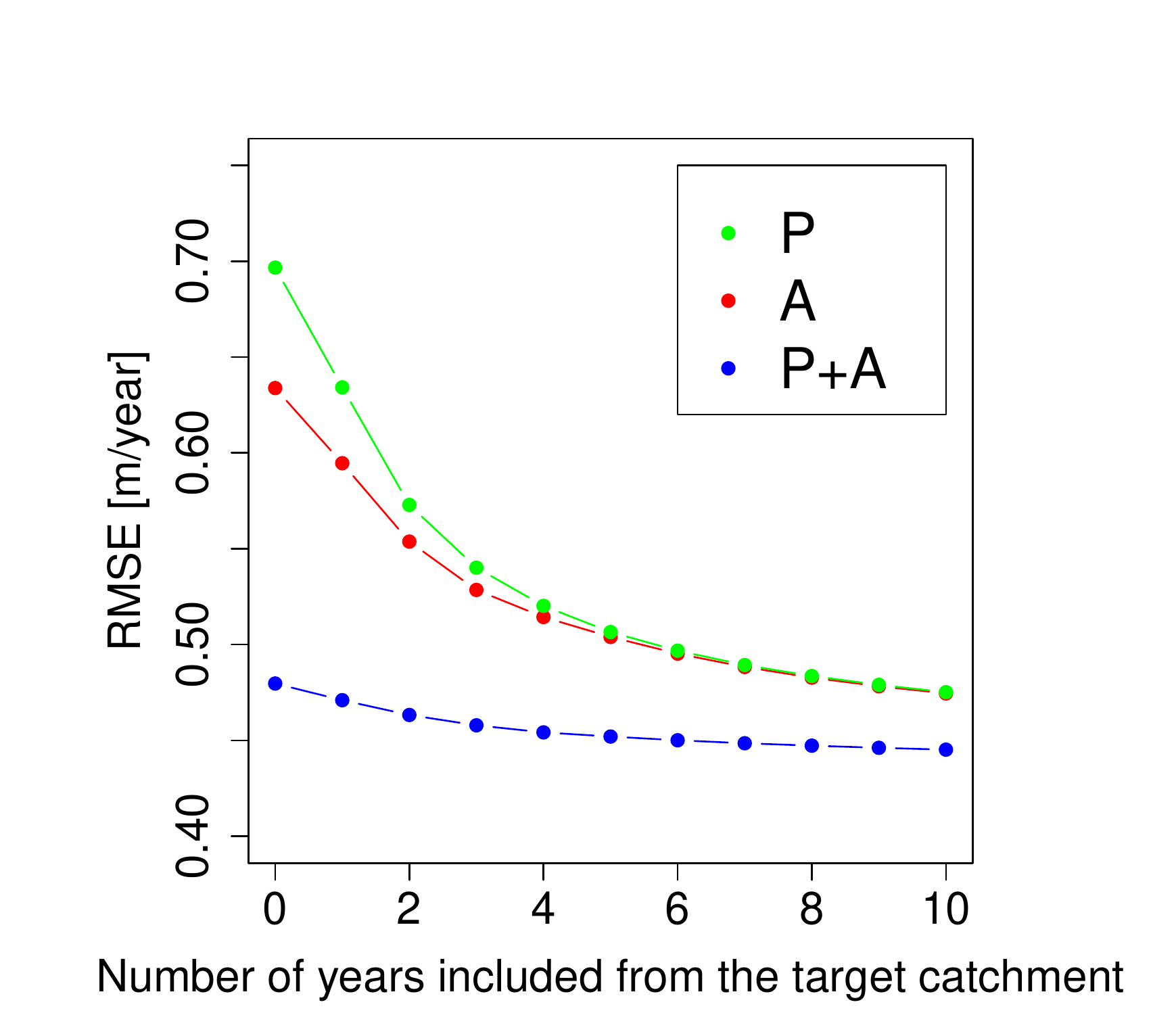}
		\caption{\small $\text{RMSE}$.}
		\label{fig:RMSE_time_real}
	\end{subfigure}
	~
	\begin{subfigure}[b]{0.4\textwidth}
		\includegraphics[page=1, trim = 5mm 1mm 0mm 1mm, clip, width=6cm]{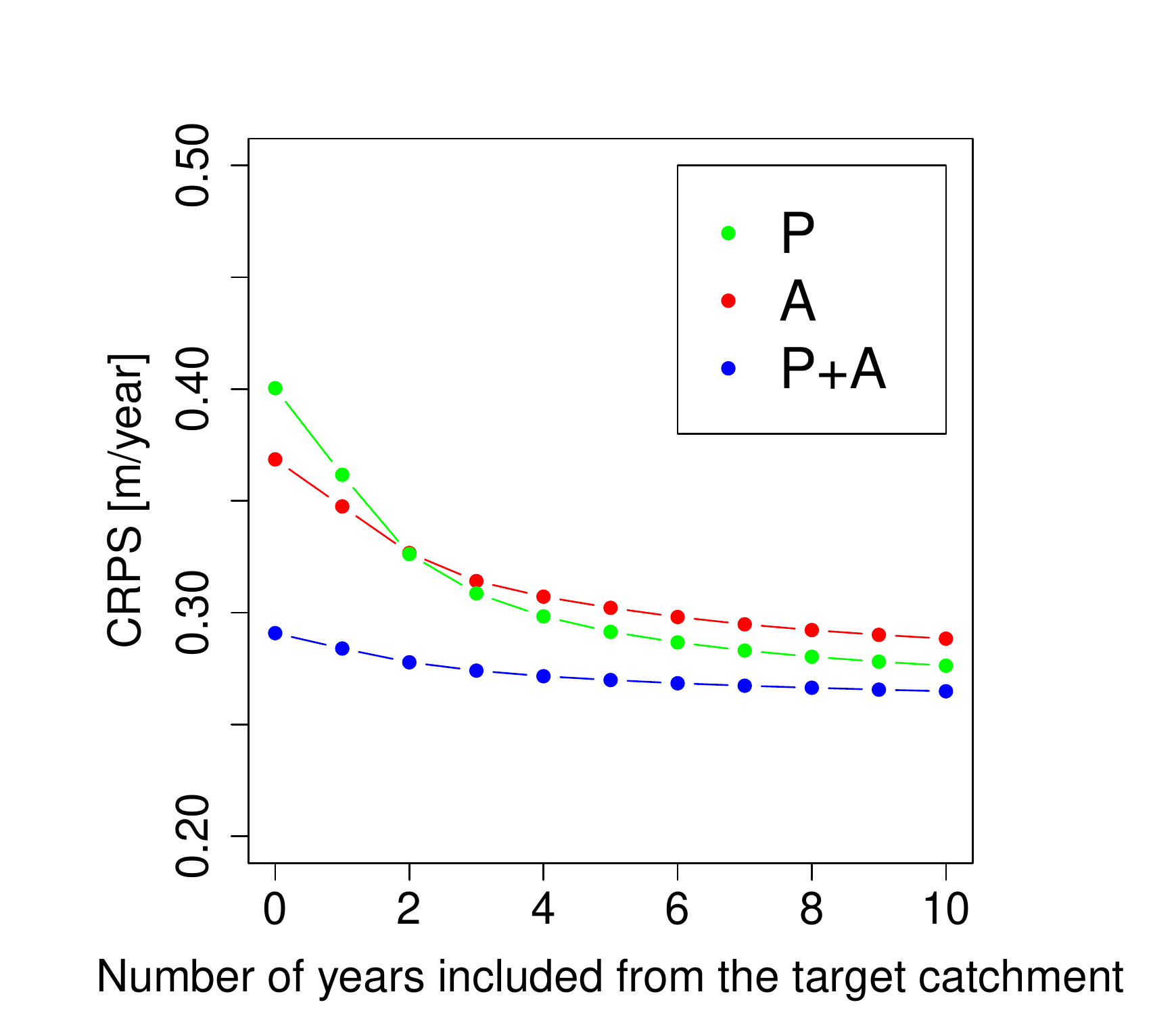}
		\caption{\small $\text{CRPS}$.} 
		\label{fig:CRPS_time_real}
	\end{subfigure}
	\caption{\small Predictive performance for future runoff (1998-2014) in catchment $\mathcal{A}_1,...,\mathcal{A}_5$ when 0-10 years of observations from the target catchment between 1988 and 1997 are included in the observation likelihood together with other observations of $P$ and/or $A$ (tests \textbf{T3u} and \textbf{T3g} ).}\label{fig:RMSE_CRPS_time_real}
\end{figure*}

In tests \textbf{T3g} and \textbf{T3u} annual runoff was predicted for unobserved future years (1998-2014) when 0-10 observations from the target catchment between 1988 and 1997 were included in the likelihood, together with observations of $P$ and/or $A$ from other locations and catchments. The resulting predictive performance is visualized in Figure \ref{fig:RMSE_CRPS_time_real}. As for the spatial predictions, $P+A$ gives the lowest RMSE and CRPS on average. For ungauged catchments (when 0 years of observations from the target catchment are included), $P$ and $A$ perform considerably worse than $P+A$. However, when we include some years of observations from the target catchment, we see a large drop in the RMSE and CRPS for $P$ and $A$. The posterior mean for a future year is given by the posterior mean of $\int_{\boldsymbol{u} \in \mathcal{A}_k}( \beta_c+c(\boldsymbol{u}))d\boldsymbol{u}$, i.e. the plots show that we get a large change in the climatic part of the model when we include information from a new location or catchment. 

This result can be understood from the results from the parameter estimation in \textbf{T1}: The posterior median of the standard deviation of the climatic  \textsc{grf} $\sigma_c$ is approximately twice as large as the median of the marginal standard deviation for the annual \textsc{grf} $\sigma_x$ for all observation designs (Table \ref{tab:parameter_estimates}). Hence, the potential value of a new data point from an unobserved location can be large, as the new observation affects the climatic part of the model that has a substantial impact on the predictions for all years under study. Furthermore, the large spatial climatic effect can also be a possible explanation for the systematic errors we saw for the spatial predictions in Figure \ref{fig:Catchment3_T2} and Figure \ref{fig:scatter_real} (\textbf{T2}). A strong climatic field $c(\boldsymbol{u})$ indicates that the same spatial runoff pattern is repeated each year, and if we fail to characterize it, systematic errors are a reasonable consequence.

\section{Simulation study}\label{sec:simulationstudy}
One of the objectives of this paper was to show how quantifying long-term spatial variability can be used as a tool for understanding the uncertainty and biases in the modeling of environmental variables. In the case study we have already suggested that a strong climatic field $c(\boldsymbol{u})$ can be an explanation for the systematic over- and underestimation we saw for some of the catchments. In the simulation study we present here, we aim to investigate this further, i.e. we explore if the over- and underestimation actually is a model property, and that it is not only caused by e.g. mismatch between the model and the runoff data in Voss. More specifically, if the true underlying process is driven by two different spatial processes, one climatic (common for all years) and one annual (different each year), can these systematic predictive biases be expected for a given catchment and set of observation locations?

In the simulation study, we explore the model properties for different values of the spatial parameters  $\rho_c , \rho_x , \sigma_c$ and $\sigma_x$. The parameters could represent different environmental variables or different study areas. By this, we aim to show what insight one can obtain about a spatio-temporal environmental variable of interest and the corresponding study area by separating climatic spatial variability from year dependent effects.

\subsection{Experimental set-up}
In the simulation study, we simulate from the model described in Section \ref{sec:models} for 9 different configurations of the range parameters $\rho_c$, $\rho_x$ and the marginal standard deviations $\sigma_c$ and $\sigma_x$. These are shown in Table \ref{tab:simparameters}. We here refer to the proportion $\sigma_c^2/(\sigma_c^2 + \sigma_x^2)$ as the \textit{climatic spatial dominance} as it represents a quantification of how large the climatic spatial effect $c(\boldsymbol{u})$ is relative to the year specific spatial effects $x_j(\boldsymbol{u})$. Note that Parameter set 1  with $\sigma_c =0.8$, $\sigma_x=0.3$, $\rho_c=20$ and $\rho_x=100$ corresponds to the posterior medians obtained for the real case study for $P+A$ (Table \ref{tab:parameter_estimates}). The other parameter sets could represent the dependency structure of another climatic variable, e.g. temperature or \textit{monthly} runoff, or the annual runoff in another part of the world.

\begin{table}[h!!!]\caption{ \small Parameters used for the simulation study. Parameter set 1 corresponds to the parameters obtained for the case study for $P+A$ in Table \ref{tab:parameter_estimates}. We refer to the proportion $\sigma_c^2/(\sigma_c^2 + \sigma_x^2)$ as the climatic spatial dominance.}\label{tab:simparameters}
	\begin{tabular}{llllll}\hline \small
		Parameter set	& $\sigma_c$ [m/year] & $\sigma_x$ [m/year] & $\rho_c$ [km] & $\rho_x$ [km] & $\sigma_c^2/(\sigma_c^2 + \sigma_x^2)$ \\
		\hline
		1 & 0.8        & 0.3        & 20       & 100      & 0.88                              \\
		2 & 0.5        & 0.5        & 20       & 100      & 0.50                   \\
		3 & 0.3        & 0.8        & 20       & 100      & 0.12                              \\
		\hline
		4 & 0.8        & 0.3        & 50       & 100      & 0.88                            \\
		5 & 0.5        & 0.5        & 50       & 100      & 0.50                           \\
		6 & 0.3        & 0.8        & 50       & 100      & 0.12                              \\ \hline
		7 & 0.8        & 0.3        & 100      & 100      &0.88                             \\
		8 & 0.3        & 0.5        & 100      & 100      & 0.50                                \\
		9 & 0.5        & 0.8        & 100      & 100      & 0.12      \\
		\hline                        
	\end{tabular}
\end{table}

The remaining two parameters are set to $\beta_c=2$ and $\tau_\beta=5$ for all experiments, i.e. similar to the posterior medians for $P+A$ in Table \ref{tab:parameter_estimates}. Furthermore, we assume that the measurement errors of the point observations are normally distributed with standard deviation $15 \%$ of the corresponding simulated value, while the measurement errors of the areal observations are normally distributed with standard deviation $3 \%$ of the corresponding simulated value. These estimates are set based on recommendations from the data provider \textsc{nve} regarding the measurement errors we typically see for precipitation and runoff.

For all 9 parameter configurations, annual runoff is simulated for the point and areas in Figure \ref{fig:data}. This way we obtain a realistic distribution of observations. In total 50 datasets were generated for each parameter set, i.e. there are 50 simulated climates  $c(\boldsymbol{u})+\beta_c$, and for each climate there are 10 replicates of the year specific component $x_j(\boldsymbol{u})+\beta_j$.

In our experiments, we predict runoff for two of the catchments in Figure \ref{fig:data}: Catchment 1 that is not nested and located relatively far from most point observations, and Catchment 4 that is nested and located in the middle of the study area with many surrounding observations. In turn, Catchment 1 or Catchment 4 is left out of the dataset, and 10 years of annual runoff (1988-1997) are predicted for the target catchment based on all point observations and the remaining areal observations from the same time period (1988-1997). That is, we use the setting $P+A$ for all simulated experiments. Furthermore, the predictions are done both when the target catchment is treated as ungauged with 0 annual runoff observation included in the likelihood, and when the target catchment is treated as partially gauged with 1 randomly drawn annual runoff observation (out of 10 years) included in the likelihood.

In order to investigate the relationship between the model parameters and prediction bias over time, we quantify bias as follows: For each of the 50 climates, we predict runoff for Catchment 1 and Catchment 4 for 10 years. Then, we compute the empirical probability that all of the 10 true (simulated) values of annual runoff are either below or above the 10 corresponding posterior medians for a specific catchment. We refer to this as the probability of \textit{systematic bias}, i.e.
\begin{equation*}
\text{Prob(Systematic bias)}=\text{Prob(All 10 simulated values are either below or above the 10 posterior medians)}.
\end{equation*}
Systematic bias was common in the case study, and can be seen for example for Catchment 5 in Figure \ref{fig:scatter_real} for P+A. We report the probability of systematic bias as \textit{one} value per parameter set, estimated based on 100 events (50 climates and 2 target catchments).

\subsection{Results from the simulation study}
We first present the overall 95 $\%$ coverage for the simulation study, based on predictions of 10 years of runoff for 2 catchments and 50 climates. These are shown in Table \ref{tab:coverage_sim}, and we find that the empirical coverages are close to 95 $\%$ for all the parameter sets in Table \ref{tab:simparameters}. If we next consider a scatter plot of the 1000 true and predicted values (not included here), the predictions are also unbiased with respect to the true runoff values. The 95 $\%$ coverages and the scatter plots confirm that the model behaves as expected asymptotically for all parameter sets.

\begin{table}[h!!] \small \caption{\small Overall 95 $\%$ coverage for the simulation study over 50 climates, 2 catchments and 10 years of predictions for $P+A$. For ungauged catchments, there are 0 observations from the target catchment in the likelihood while for partially gauged catchments there is 1 annual observation available from the target catchment.} \label{tab:coverage_sim}
	\begin{tabular}{llllllllll}\hline
		Parameter set.              & 1    & 2    & 3    & 4    & 5    & 6    & 7    & 8    & 9    \\\hline 
		Ungauged catchments         & 0.96 & 0.96 & 0.94 & 0.94 & 0.98 & 0.96 & 0.96 & 0.96 & 0.96 \\
		Partially gauged catchments & 0.93 & 0.93 & 0.94 & 0.95 & 0.95 & 0.96 & 0.96 & 0.96 & 0.95\\
		\hline
	\end{tabular}
\end{table}

\begin{figure}
	\centering
	\begin{subfigure}[b]{0.45\textwidth}
		\includegraphics[width=0.9\textwidth]{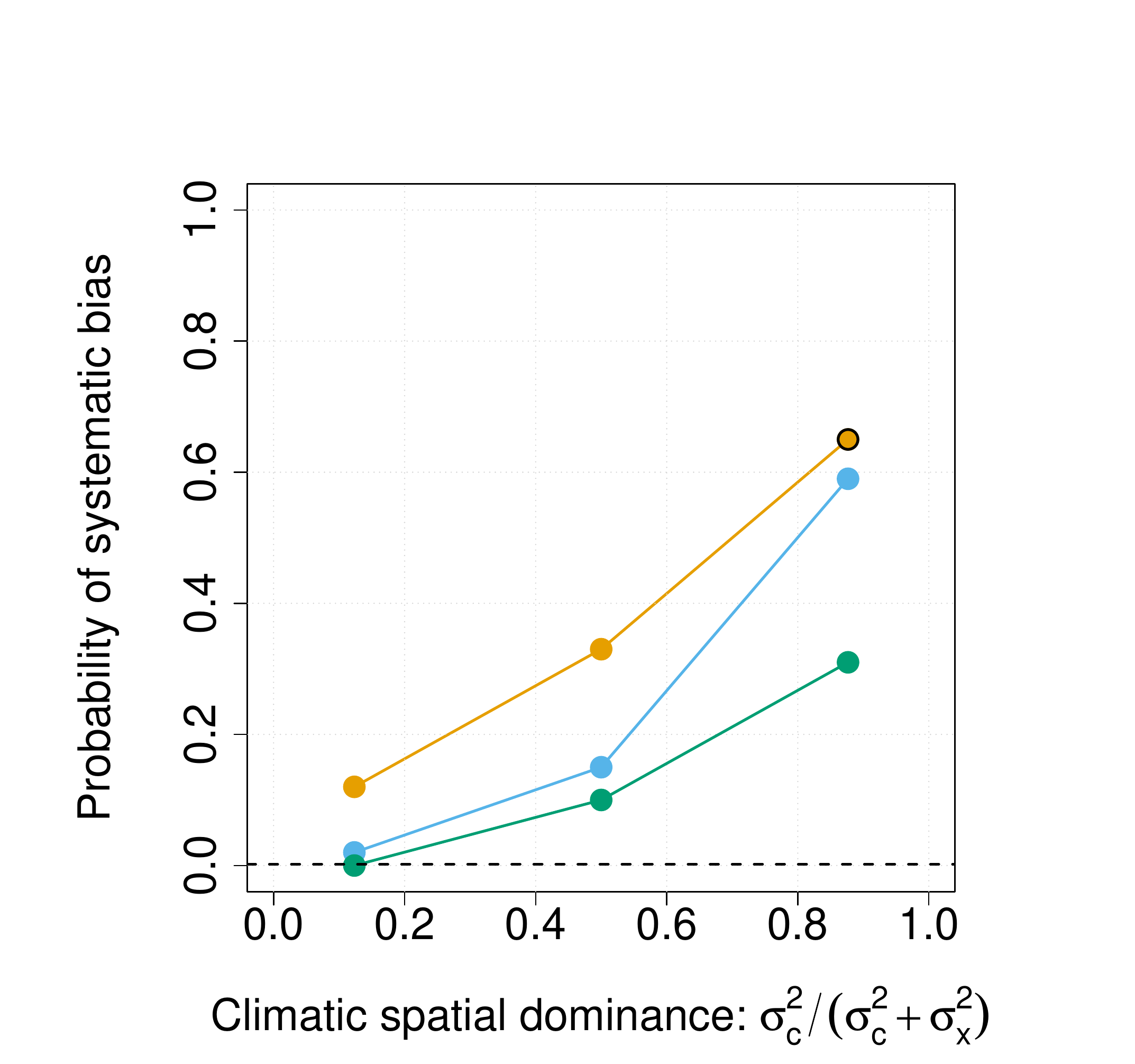}
		\caption{\small 0 observations from the target catchment.}
		\label{fig:bias0}
	\end{subfigure}
	~ 
	\begin{subfigure}[b]{0.45\textwidth}
		\includegraphics[width=0.9\textwidth]{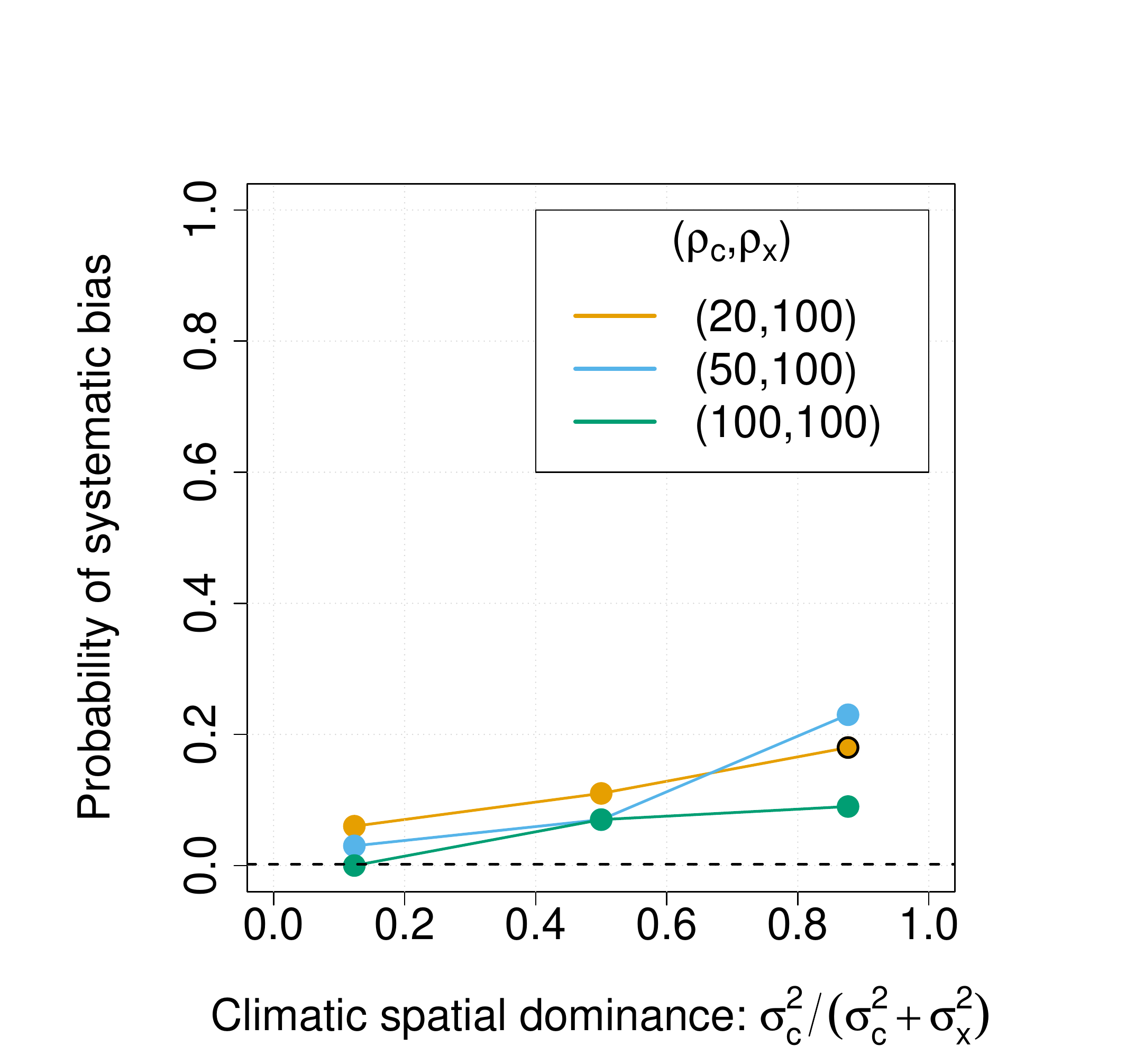}
		\caption{\small 1 observation from the target catchment.}
		\label{fig:bias1}
	\end{subfigure}
	~ 
	\caption{ \small The empirical probability that all of the 10 true annual values are either above or below the posterior median value for Catchment 1 and Catchment 4 over 50 climates for ungauged catchments (Figure \ref{fig:bias0}) and partially gauged catchments (Figure \ref{fig:bias1}). The black circle corresponds to the parameter values we have for the case study from Voss. The black dashed line is the theoretical probability that all the observed values are above or below the posterior median when studying a process that actually is independent over years (2 $\cdot$ $0.5^{10} = 0.2 \%$ ). This is included as a reference.  }\label{fig:bias}
\end{figure}

Next, Figure \ref{fig:bias0} shows an visualization of the systematic bias obtained for the simulation study when the target catchments (Catchment 1 and Catchment 4) are treated as ungauged. Recall that systematic bias here is measured as the probability that all 10 true annual runoff values are either below or above the corresponding predicted value for a specific climate and catchment. We see a clear relationship between this bias and the climatic spatial dominance given by the proportion $\sigma_c^2/(\sigma_c^2 + \sigma_x^2)$: 
When annual spatial effects dominate over climatic spatial effects and $\sigma_c  \ll \sigma_x$, the probability of systematic bias is close to zero (around 0.2 $\%$). However, when most of the spatial variability is due to the climate ($\sigma_c\gg\sigma_x$), this probability increases to 30-65$\%$ depending on the values of the range parameters $\rho_c$ and $\rho_x$. For the parameters corresponding to the Norwegian case study, the probability of systematic bias was 65 $\%$. Hence, systematic errors like we saw for e.g. Catchment 5 (P+A) in Figure \ref{fig:scatter_real}, can be expected quite often for these parameter values. Figure \ref{fig:bias0} also shows that the probability of systematic bias is largest when the climatic range $\rho_c$ is low, i.e. when the information gain from the neighboring catchments is low.

From a statistical point of view, the above results are intuitive: If most of the spatial variability can be explained by climatic conditions, there are large dependencies between years. Either we typically perform accurate predictions all years, or poor predictions all years. Considering all ungauged catchments in Norway, we can expect that $95 \%$ of the true runoff values are inside the corresponding $95 \%$ posterior prediction intervals on average (Table \ref{tab:coverage_sim}), but if we consider predictions for individual catchments over time, a large proportion of the predictions will be biased in one direction or the other (Figure \ref{fig:bias0}). The simulation study shows that the systematic bias we obtained for the case study are not necessarily a result of mismatch between the data and the fitted model, but can indeed be a result of the strong climate around Voss ($\sigma_c\gg\sigma_x$).

So far we have considered the probability of systematic bias when there are no data from the target catchments available. Next, in Figure \ref{fig:bias1}, we present the probability of systematic bias when there is one annual observation included in the likelihood. For $\sigma_c\ll \sigma_x$, i.e. when  $\sigma_c^2/(\sigma_c^2+\sigma_x^2)$ is close to zero, we see that the probability of systematic bias in general is low for both ungauged catchments (Figure \ref{fig:bias0}) and partially gauged catchments (Figure \ref{fig:bias1}). For this scenario, a new data point from the target catchment don't have a considerable impact on the probability of systematic bias. However, if $\sigma_c \gg \sigma_x$ as in Voss, we find that the extra data point on average leads to a large reduction in the systematic bias probability in Figure \ref{fig:bias1} compared to the systematic bias probability we saw for the ungauged catchments in Figure \ref{fig:bias0}. This is found for all combinations of $\rho_c$ and $\rho_x$, but the tendency is strongest if $\rho_c \ll \rho_x$ as in Voss. The results in Figure \ref{fig:bias} are thus comparable to the results in Figure \ref{fig:RMSE_CRPS_time_real} for the case study, and illustrate the potential value of data from a new location for different parameter values. 

\section{Discussion}\label{sec:discussion}
In this paper we have presented a model for annual runoff that consistently combines data of different spatial support. The suggested model is a geostatistical model with two spatial effects: A climatic long-term effect and a year dependent  effect that describes the annual discrepancy from the climate. The model was used to estimate mean annual runoff in the Voss area in Norway.

The main focus of the study was on exploring how the combination of point and nested areal observation affects runoff predictions, to demonstrate that our model has mass-conserving properties and to show how quantifying long-term spatial variability can be used as a tool for understanding biases in environmental modeling and for exploiting short records of data. There are three key findings: 1) On average we benefit from including all available observations in the likelihood, both point and areal data. $P+A$ performed better than $P$ and $A$ in terms of RMSE, CRPS and the coverage of the 95 $\%$ posterior prediction intervals in our case study. $P+A$ also performed better than the referenced method Top-Kriging that only supports areal observations. 2) The suggested model that combines point and areal observations is particularly suitable for modeling the nested structure of catchments. The case study showed that the model was able to identify Catchment 3 as a wetter catchment than any of the surrounding catchments and precipitation stations. This was a consequence of using information from two overlapping catchments to constrain and distribute the annual runoff correctly. The interaction between the point and nested areal observations gives a geostatistical model that does more than smoothing, and this represents a main difference from e.g. Top-Kriging that doesn't constrain the predicted runoff. 3) How dominating climatic spatial effects are compared to annual spatial effects has a large influence on the predictability of runoff. If most of the spatial variability can be explained by long-term (climatic) weather patterns and processes, systematic biases for a location over time can be expected as long as the same observation design is used.

The fact that $P+A$ performed better than $A$ for most catchments around Voss, indicates that the point and areal observations of runoff were sufficiently compatible for most catchments, i.e. that evaporation subtracted from precipitation was a valid approximation of point runoff. This interpretation of point runoff is reasonable in areas like Voss where the annual precipitation is considerably larger than the annual evaporation. The evaporation data are uncertain and should not make a large impact on the resulting predictions. In many areas of the world, the observed annual evaporation is more than 50 $\%$ of the annual observed precipitation. In such areas, our framework could provide negative point observations and results that are hard to interpret. Negative runoff can in general be a problem in our Gaussian model. Log transforming the data is a solution if considering only point data ($P$), but is not an option when modeling areal data ($A$ and $P+A$) because the log transformation does not work well with the linear aggregation in Equation \eqref{eq:runoffbasis}. For areas with observed values close to zero, extra caution should therefore be taken regarding negative, non-physical results. To avoid negative predictions it is also important to make sure that the mesh used in the \textsc{spde} approach (Figure \ref{fig:catchmesh}) is fine enough to capture the rapid spatial variability in the study area.

Precipitation observations are often avoided as an information source when performing interpolation of runoff in hydrological applications, but the results presented here show that the point observations can contain valuable information when used together with areal observations. At least in data sparse areas with few streamflow observations. However, there is still room for improvement in the compatibility between the two observation types: The observation designs including only point observations $P$ provided a clear underestimation of annual runoff for most catchments in the case study. It was also seen that the spatial field provided by the precipitation observations ($P$) was smoother than the spatial field provided by the runoff observations ($A$) in Figure \ref{fig:future_eta_mean}, which is a typical result: The increase in spatial variability from precipitation to runoff is mainly explained by small scale variability introduced by soil and vegetation \citep{smoothfield}. Consequently, if the point data are allowed to dominate over the areal data, the point data can cause a runoff field that is too smooth, which affects both the posterior mean and the posterior standard deviation disadvantageously.

Furthermore, it is worth mentioning that all of the available precipitation gauges are located at a lower elevation than the mean elevation of the five catchments in the dataset. This is a common problem. Precipitation gauges are often located at low elevations, close to settlements where the gauges are easy to maintain. It is known that the amount of precipitation typically increases with elevation. There is therefore a lack of information about precipitation at high elevations in the data. Adding the fact that the precipitation gauges often fail to catch a large proportion of the precipitation, in particular when it comes as snow and it is windy \citep{kochen1}, essential information about the precipitation and runoff field could be lost. To solve the compatibility issues, elevation was considered as a covariate in a preliminary study \citep{Jorid}, but this did not lead to significant improvements, and the results are not included here. Another option could be a preferential sampling approach where we assume that the locations of the precipitation gauges are distributed according to a log-Gaussian Cox process that depends on the response variable, here through elevation implicitly \citep{diggle}.
 
Elevation is also known to be a factor that affects the spatial dependency structure of precipitation, and Voss is a mountainous area. The spatial range is typically larger in lowlands and decreases with elevation. A non-stationary model similar to the one presented in \cite{Rikke1} with a range and a marginal variance that changes with elevation could be considered. This can easily be implemented within the \textsc{inla}-\textsc{spde} framework. However, in this case the dataset is small and the complexity of the spatial variability large. We also have a model with only one replication of the climatic spatial effect which was the dominating spatial component. A non-stationary model would probably be too complicated and lead to identifiability issues \citep{Rikke2}.

Regardless of the increased complexity in an extended model, it is reasonable to believe that an accurate representation of the climatic conditions at a target location is crucial when predicting annual runoff and other climate related variables. In the simulation study, we demonstrated how systematic under- and overestimation of a target variable can be expected over time when we fail to characterize the underlying climate in areas where the climatic spatial field's marginal standard deviation $\sigma_c$ is large relatively to the other model standard deviations. We also found a clear relationship between the model parameters of the suggested model, and systematic prediction bias over time. This shows that the two field model (and its parameters) can contribute with useful insight about the properties of a study area and/or an environmental variable of interest.

In spite of the large biases documented for annual runoff predictions in this article, a dominating climate also gives opportunities. In this article a model with a climatic component was suggested. The climatic component included a spatial effect that was common for all years of observations. This component made it relatively simple to exploit short records of data, and the runoff predictions could easily be improved by including a few observations from the target catchments. Time series from several years are not needed because one or two observations from a new catchment updates the climatic component that has a large impact to the final model if $\sigma_c$ dominates over the other model variances. Here, we again note how the model parameters can contribute with useful information about the study area and/or the environmental variable of interest: The potential gain of collecting a new data point from a new location, i.e. a short record, can be indicated from the spatial parameters, in particular from the proportion $\sigma_c^2/(\sigma_c^2 + \sigma_x^2).$  

The ability to exploit short records is another main benefit of the suggested model over existing spatial models used for runoff interpolation, like e.g. Top-Kriging. For practitioners, a model with the described properties can be useful in situations where there exist one or few observations from a catchment of interest. Short duration runoff observations are quite common in hydrological datasets, e.g. from planned short duration missions for water resources assessments, or from gauging stations that are closed after a revision of the gauging network. Large infrastructure projects measuring a few years of annual runoff for a relevant catchment is also achievable.

\bigskip
	

\bibliographystyle{plainnat}
\bibliography{masterbib2}

\end{document}